\documentclass[12pt]{article}

\usepackage[a4paper,margin=1in]{geometry}
\usepackage{amsmath,amssymb}
\usepackage{booktabs}
\usepackage{graphicx}
\usepackage{float}
\usepackage{longtable}
\usepackage{array}
\usepackage{caption}
\usepackage{subcaption}
\usepackage[colorlinks=true,linkcolor=blue,citecolor=blue,urlcolor=blue]{hyperref}
\graphicspath{{figures/}}

\newcommand{\pct}[1]{#1\%}
\newcommand{\code}[1]{\texttt{\detokenize{#1}}}
\newcommand{\pathcode}[1]{\begingroup\scriptsize\url{#1}\endgroup}

\title{\bfseries Continuous Cash-Overlay Filters for a Static Growth--Defensive Risk Sleeve\\
\large Direct Slow-Tail Compensation, Fixed V-Shape Crash Brakes, Walk-Forward Validation, and Max-Cash Combination}
\author{Zheli Xiong\\
\normalsize Corresponding author: Zheli Xiong (zlxiong@mail.ustc.edu.cn)\\
\normalsize Code repository: \url{https://github.com/shaun19920309/gd-cash-overlay-filters}}
\date{}

\begin{document}
\maketitle

\begin{abstract}
This paper studies a modular cash-overlay rule for allocating between a fixed growth--defensive risky sleeve $R$ and interest-bearing cash $C$. The risky sleeve is a static 50/50 combination of equal-weight growth/technology and defensive income/value ETF baskets; the target is future $R-C$ return with the cash leg earning the contemporaneous cash rate. Two independent filters are tested. The slow-tail filter maps continuous compensation, rate-headwind, risk-premium-compression, and rate-path-stress states into a cash weight with a 30\% material-trade gate. The V-shape filter is a fast crash brake based on continuous VIX, rate, credit, drawdown, and re-entry states. A fixed max-cash layer then uses the larger cash weight requested by either filter each day. On the 2017--2026 common window, the selected max-cash combination earns an 18.83\% CAGR versus 16.62\% for 100\% $R$ and reduces maximum drawdown from -33.59\% to -18.05\%. In the main walk-forward OOS window, the expanding combination earns 19.35\% versus 17.59\% for 100\% $R$, with maximum drawdown of -22.05\% versus -33.59\%; the rolling version earns 18.50\% with the same -22.05\% drawdown. Post-2022 tests show lower drawdown but lower CAGR during a strong risky-sleeve rebound. The results support modular cash overlays as drawdown-control tools rather than standalone return-enhancement claims; fully real-time variable re-screening and multiple-testing-adjusted inference remain future work.
\end{abstract}

\noindent\textbf{Keywords:} dynamic asset allocation; cash overlay; crash protection; VIX; interest rates; credit stress; walk-forward validation; drawdown control.

\section{Introduction}

This paper asks whether a static growth--defensive risky sleeve should sometimes be replaced by interest-bearing cash. The problem is deliberately narrower than a full asset-allocation model. The risky sleeve $R$ is fixed, transparent, and independent of any dynamic growth--defensive timing rule. The overlay only decides how much of this risky sleeve to hold versus cash.

The research design is modular, but the two modules do not change equally in this version. The V-shape branch is retained as a fixed crash brake for fast drawdowns. The slow-tail branch is rebuilt as a direct compensation score rather than a historical analogue estimator. The two filters are then combined with a conservative max-cash rule. This design avoids a single broad regime classifier and instead treats persistent compensation deterioration and fast crash risk as different alarms.

The paper follows an empirical strategy-study structure. It first defines the static risky sleeve and the cash benchmark. It then documents the statistical screening logic, the two policy rules, the parameter grids, full-sample selected results, walk-forward validation, event diagnostics, and the final max-cash combination. Appendix tables preserve the main screening and robustness outputs.

\section{Related Literature}

The study sits between return-predictability research, parametric portfolio policy design, and practical crash-risk management. The predictive-return literature warns that variables with in-sample explanatory power often fail in out-of-sample settings, especially when candidate variables are numerous and labels overlap \cite{goyal2008comprehensive,campbell2008predicting}. This motivates the paper's separation between diagnostic screening and trading evidence: Newey--West regressions, non-overlapping labels, and multi-horizon consistency are used to organize candidate state variables, while walk-forward portfolio tests decide whether a component is useful as a cash overlay.

The policy design is closest in spirit to parametric allocation rules. Brandt, Santa-Clara, and Valkanov \cite{brandt2009parametric} show how portfolio weights can be linked directly to state variables. Instead of estimating an unconstrained mean-variance optimizer, this paper maps economically interpretable state variables into a continuous cash weight. This keeps the action space narrow: the risky sleeve is fixed, and the policy can only reduce or restore exposure to that sleeve. The material cash gate of 30\% is not a predictive threshold. It is an implementation rule that says a computed cash allocation must be large enough to matter before a trade is taken.

The robustness discussion follows the data-snooping literature. Full-sample screening and policy construction are useful for discovering economically coherent mechanisms, but they do not by themselves establish a final anomaly claim. White's Reality Check, Hansen's SPA test, and the Deflated Sharpe Ratio are therefore treated as future confirmation tools rather than as claims already completed in this version \cite{white2000reality,hansen2005superior,bailey2014probability}. The validation discipline follows the companion growth--defensive style-timing study \cite{xiong2026continuous}: first build a transparent empirical mechanism, then test whether it survives increasingly strict validation layers.

\section{Static Risky Sleeve and Cash Return}

Let $G_t$ denote the equal-weight return of the growth/technology basket \{QQQ, XLK, VGT, SPYG, VUG\}. Let $D_t$ denote the equal-weight return of the defensive income/value basket \{SCHD, VYM, VTV, FDVV, COWZ\}. The risky sleeve is
\begin{equation}
R^R_t = 0.5G_t + 0.5D_t.
\end{equation}
Because each basket contains five ETFs, this is equivalent to a 10-ETF equal-weight risky sleeve. The cash sleeve return $R^C_t$ is constructed from the contemporaneous cash yield. The target excess return for screening is
\begin{equation}
R^{R-C}_t = R^R_t - R^C_t.
\end{equation}
This construction is intentionally independent of the dynamic growth--defensive style-allocation policy studied in \cite{xiong2026continuous}. That companion policy may later allocate within the risky sleeve between growth and defensive assets, but the cash-overlay experiments in this paper do not use its scores, states, or weights. Throughout the paper, ``100\% $R$'' means the static 50/50 $G/D$ risky sleeve, not a dynamic style-timing policy.

\begin{table}[H]
\centering
\caption{Data Coverage and Portfolio Definitions}
\label{tab:data-coverage}
\scriptsize
\begin{tabular}{p{3.3cm}p{5.4cm}p{5.4cm}}
\toprule
Object & Definition & Sample or Role\\
\midrule
Growth basket $G$ & Equal-weight QQQ, XLK, VGT, SPYG, VUG & Technology and growth equity exposure.\\
Defensive basket $D$ & Equal-weight SCHD, VYM, VTV, FDVV, COWZ & Dividend, value, quality, and cash-flow tilted equity exposure.\\
Risky sleeve $R$ & $0.5G+0.5D$ & Static 10-ETF equal-weight risky pool used in every cash-overlay comparison.\\
Cash sleeve $C$ & Contemporaneous cash/risk-free return & Interest-bearing cash benchmark; cash yield enters both returns and state variables.\\
Feature panel & 2017-06-28--2026-04-30, 2222 trading days & Shared state panel used by slow-tail and V-shape modules.\\
Main OOS window & 2018-06-28 requested start, realized from first valid 63-day test block & Walk-forward parameter re-selection.\\
Post-2022 OOS window & 2022-01-03--2026-04-30 & Stress test for the slow-tail regime and the post-rate-hike environment.\\
\bottomrule
\end{tabular}
\end{table}

The final common-window comparisons apply a 10bp one-way transaction cost unless otherwise stated. Turnover is measured as the one-way change in the risky weight, $|w^R_t-w^R_{t-1}|$, whenever selected component weights and max-cash combined weights are recomputed on the common daily panel. Component-native diagnostic tables retain the archived module convention used by the corresponding component script. This matters mainly for the V-shape standalone tables, whose native turnover counts both risky-weight and cash-weight changes; the max-cash section therefore reports a common-window V-shape-only recomputation before combining the modules.

\subsection{Raw State Variables}

The state panel is built from observable prices, yields, and spread proxies. ETF returns use adjusted daily close data. Treasury and credit-yield variables enter in percentage-point units before standardization. VIX variables enter either as rolling percentiles or percentage changes, and credit-risk proxies use both cash-bond yield spreads and ETF relative returns. The risky-sleeve variables are rebuilt from the static $R$ definition in this paper, so no companion style-timing score, state, or dynamic $G/D$ weight enters the cash-overlay component panel.

\begin{table}[H]
\centering
\caption{Raw State Variables Used in Cash-Overlay Components}
\label{tab:raw-state-variables}
\scriptsize
\begin{tabular}{p{3.2cm}p{5.5cm}p{5.4cm}}
\toprule
Variable & Construction & Component Role\\
\midrule
$R^R_t$ & $0.5G_t+0.5D_t$, where $G$ and $D$ are equal-weight ETF baskets. & Risky-sleeve return used in both strategy returns and all $R$-dependent state variables.\\
$R^C_t$ & Daily interest-bearing cash return from the contemporaneous cash/risk-free series. & Cash benchmark return; $R-C$ is the screening target.\\
Cash yield & Annualized cash yield, with 21-day change. & Cash compensation level and cash-rising pressure.\\
10-year Treasury yield & 21-day change in TNX. & Rate headwind in slow-tail; rate relief in V-shape after sign reversal.\\
10-year minus 3-month curve & TNX minus IRX. & Curve-inversion diagnostic in the main-effect screen.\\
$R$ drawdown & Cumulative $R$ wealth divided by its running peak minus one. & Drawdown depth and oversold state.\\
$R$ trailing return & 10-, 21-, 63-, 126-, and 252-day cumulative $R$ return, with the 10-day version used only in V-shape crash-loss interactions. & Risky-sleeve strength, underwater state, and short-horizon crash loss.\\
$R$ volatility & 63-day realized volatility of $R$, annualized by $\sqrt{252}$. & Low-volatility compensation state.\\
VIX level & 756-day rolling percentile of VIX. & High/low volatility state and V-shape level core.\\
VIX changes & 21-day VIX change for slow-tail; 10-day VIX change and 5-day VIX relief for V-shape. & Volatility spike and volatility relief.\\
HYG--SHY relative return & 21-day HYG return minus 21-day SHY return; 5-day version is used for V-shape credit appetite. & Credit risk-off and credit appetite.\\
BAA--AAA spread & Moody's BAA yield minus AAA yield, and its 21-day change. & Credit widening and panic interaction.\\
\bottomrule
\end{tabular}
\end{table}

All continuous states are transformed with expanding standardization:
\begin{equation}
z_t(x)=\frac{x_t-\mu^{exp}_t(x)}{\sigma^{exp}_t(x)}.
\end{equation}
For direct slow-tail components, the expanding window uses a 252-trading-day warmup and a one-day lag before a standardized value is used. V-shape primitives use the same expanding-standardization idea, with the candidate score layer standardized again before the deployable brake score is formed. The current V-shape branch is retained from the fixed material-brake version; the methodological update in this paper is the direct slow-tail compensation score.

\section{Statistical Screening and Validation Design}

Directional intensities use smooth transformations such as $softplus(x)=\log(1+\exp(x))$. The retained filters do not use binary feature truncation. The only hard trading gate is economic materiality: a cash move is implemented only when the continuously computed raw cash weight exceeds 30\%.

Candidate variables are screened against future $R-C$ returns at 21, 63, and 126 trading-day horizons. Because labels overlap, the primary regression diagnostic is the OLS coefficient with Newey--West/HAC standard errors. The screen also reports non-overlapping 63-day regressions, multi-horizon consistency, economic sign, and coverage. These tests are used to organize candidate components. Trading evidence is evaluated separately through full-sample grid selection and walk-forward parameter re-selection.

For each strategy, the full-sample grid is a descriptive fitted result. In OOS tests, each 63-trading-day test block re-selects the complete parameter vector using either all prior data (expanding) or the most recent 756 trading days (rolling). Thus, OOS return series are not fixed-parameter backtests. Appendix Tables~\ref{tab:oos-complete} and~\ref{tab:wf-selection} report the complete OOS return matrix and the block-level parameter-selection diagnostics.

\section{Baseline Annual Diagnostics}

Before evaluating timing rules, it is important to separate three facts. First, the static risky sleeve has a high unconditional return over the sample, so a cash overlay must preserve most upside participation. Second, cash becomes economically relevant only after rate normalization; it cannot explain the 2020 crash-protection result. Third, the two modules add value in different years: the V-shape module matters most in 2020 and 2025, while the slow-tail module matters most in 2022.

The annual decomposition makes the modular nature of the overlay visible. In Table~\ref{tab:annual-baseline}, the combined policy is not uniformly more aggressive than its components. It matches the V-shape module in 2020, matches the slow-tail module in 2022, and accepts modest opportunity cost in calm years when a component remains partially in cash.

\begin{table}[H]
\centering
\caption{Annual Return / Maximum Drawdown by Policy, Selected-Weight Common Window}
\label{tab:annual-baseline}
\begingroup
\tiny
\setlength{\tabcolsep}{2pt}
\resizebox{\textwidth}{!}{%
\begin{tabular}{lrrrrrr}
\toprule
Year & 100\% $R$ & 100\% $C$ & Slow-tail & V-shape & Max-cash & Avg. Max-cash Cash\\
\midrule
2018 & \pct{-3.02} / \pct{-19.71} & \pct{2.54} / \pct{0.00} & \pct{-3.02} / \pct{-19.71} & \pct{-2.31} / \pct{-16.77} & \pct{-2.31} / \pct{-16.77} & \pct{7.68}\\
2019 & \pct{32.90} / \pct{-8.03} & \pct{2.55} / \pct{0.00} & \pct{32.90} / \pct{-8.03} & \pct{32.39} / \pct{-8.03} & \pct{32.39} / \pct{-8.03} & \pct{0.57}\\
2020 & \pct{23.82} / \pct{-33.59} & \pct{0.62} / \pct{0.00} & \pct{23.82} / \pct{-33.59} & \pct{44.55} / \pct{-15.11} & \pct{44.55} / \pct{-15.11} & \pct{7.64}\\
2021 & \pct{31.04} / \pct{-4.92} & \pct{0.00} / \pct{0.00} & \pct{30.66} / \pct{-4.92} & \pct{29.36} / \pct{-4.92} & \pct{28.99} / \pct{-4.92} & \pct{1.30}\\
2022 & \pct{-17.06} / \pct{-23.78} & \pct{1.29} / \pct{0.00} & \pct{-8.95} / \pct{-18.05} & \pct{-17.06} / \pct{-23.77} & \pct{-8.95} / \pct{-18.05} & \pct{15.60}\\
2023 & \pct{28.14} / \pct{-9.73} & \pct{5.13} / \pct{0.00} & \pct{27.80} / \pct{-9.73} & \pct{28.14} / \pct{-9.73} & \pct{27.80} / \pct{-9.73} & \pct{0.56}\\
2024 & \pct{22.51} / \pct{-8.88} & \pct{5.17} / \pct{0.00} & \pct{22.51} / \pct{-8.88} & \pct{18.60} / \pct{-7.93} & \pct{18.60} / \pct{-7.93} & \pct{3.28}\\
2025 & \pct{17.31} / \pct{-19.66} & \pct{5.13} / \pct{0.00} & \pct{17.31} / \pct{-19.66} & \pct{15.79} / \pct{-14.64} & \pct{15.79} / \pct{-14.64} & \pct{7.02}\\
2026 & \pct{8.61} / \pct{-7.95} & \pct{1.03} / \pct{0.00} & \pct{7.32} / \pct{-7.81} & \pct{6.64} / \pct{-7.86} & \pct{6.80} / \pct{-7.73} & \pct{7.97}\\
\bottomrule
\end{tabular}
}
\endgroup
\end{table}

\section{Policy Grids and Selection Objectives}

Both modules select score-construction parameters and position-mapping parameters. The direct slow-tail grid has 216 configurations:
\begin{equation}
216
=2\;\alpha_{comp} \times 2\;\lambda_{rate} \times 2\;\lambda_{path}
\times 3\;max\ cash \times 3\;\tau \times 3\;convexities .
\end{equation}
The V-shape grid has 216 configurations:
\begin{equation}
216
=2\;\alpha_{vix} \times 2\;\lambda_{rate} \times 2\;\lambda_{credit}
\times 3\;max\ cash \times 3\;\tau \times 3\;convexities .
\end{equation}
The material threshold is fixed at 30\% in both grids. Entry to cash is immediate for V-shape and economically immediate for slow-tail because the target cash weight is used once it is material. Re-entry to $R$ is smoothed with $\eta_{exit}=0.25$.

The grid objective is deliberately multi-criteria. A pure CAGR sort would choose overly aggressive rules that fail the risk-management purpose of the cash overlay. A pure drawdown sort would over-allocate to cash. The selection score ranks configurations by return, drawdown, Calmar, event performance, turnover, and average cash use. The exact weights differ by module because the slow-tail module is a compensation filter and the V-shape module is a crash brake, but both penalize high turnover and unnecessary cash exposure.

\begin{table}[H]
\centering
\caption{Parameter Grid Design}
\label{tab:grid-design}
\scriptsize
\begin{tabular}{p{3.2cm}p{5.0cm}p{6.0cm}}
\toprule
Module & Grid Parameters & Interpretation\\
\midrule
Slow-tail & $\alpha_{comp}\in$ \{0.50,0.67\}; $\lambda_{rate}\in$ \{0.25,0.50\}; $\lambda_{path}\in$ \{0.25,0.50\}; max cash $\in$ \{0.50,0.75,1.00\}; cash $\tau\in$ \{0.75,1.00,1.25\}; convexity $\in$ \{2,3,4\}. & Chooses the weight on cash-rising pressure, compressed-risk-premium components, and rate-path stress, then maps the direct score into a cash weight.\\
V-shape & $\alpha_{vix}\in$ \{0.50,0.67\}; $\lambda_{rate}\in$ \{0.25,0.50\}; $\lambda_{credit}\in$ \{0,0.25\}; max cash $\in$ \{0.50,0.75,1.00\}; brake $\tau\in$ \{0.50,0.75,1.00\}; convexity $\in$ \{3,4,5\}. & Chooses the mix of VIX level/spike, rate panic, and credit panic, then maps the crash-brake score into a cash weight.\\
Max-cash combination & No new fitted parameter. & Uses $w^C_t=\max(w^{C,slow}_t,w^{C,vshape}_t)$; when both modules are evaluated OOS, the combination uses their OOS-selected weights, not full-sample component weights.\\
\bottomrule
\end{tabular}
\end{table}

\begin{table}[H]
\centering
\caption{Selected Full-Sample Results, 10bp Cost}
\label{tab:selected-full}
\begingroup
\tiny
\setlength{\tabcolsep}{2pt}
\resizebox{\textwidth}{!}{%
\begin{tabular}{llrrrrrrr}
\toprule
Policy & Window & CAGR & Sharpe & Max DD & Ann. Turnover & Avg. Cash & 100\% $R$ CAGR & 100\% $R$ Max DD\\
\midrule
Slow-tail & 2018-06-28--2026-04-30 & \pct{17.83} & 0.94 & \pct{-33.59} & \pct{108.36} & \pct{2.19} & \pct{16.69} & \pct{-33.59}\\
V-shape & 2017-07-19--2026-04-30 & \pct{17.32} & 1.07 & \pct{-23.77} & \pct{532.41} & \pct{3.65} & \pct{16.62} & \pct{-33.59}\\
Max-cash combination & 2017-07-19--2026-04-30 & \pct{18.83} & 1.18 & \pct{-18.05} & \pct{355.82} & \pct{5.56} & \pct{16.62} & \pct{-33.59}\\
\bottomrule
\end{tabular}
}
\endgroup
\end{table}

The selected-weight comparison is best read as a fitted economic map rather than as the main OOS claim. Table~\ref{tab:selected-full} shows the role each module is designed to play before moving to walk-forward tests. The redesigned slow-tail filter raises CAGR modestly and improves 2022, but it leaves the full-sample maximum drawdown unchanged because it is not a fast-crash tool. The fixed V-shape branch gives a smaller CAGR gain but materially reduces maximum drawdown. The max-cash combination improves both dimensions because it inherits slow-tail protection in 2022 and V-shape protection in fast crashes. Later combination tables recompute the same selected component weights on a common panel with the final one-way turnover convention, so the V-shape-only row in Table~\ref{tab:combo-full} is intentionally not identical to the standalone V-shape row in this summary table.

\section{Slow-Tail Compensation Filter}

\subsection{Policy Construction}

The slow-tail filter is the part of the paper that changes materially in the current experiment. The previous historical-analogue construction is replaced by a direct continuous compensation score. The goal is still the same: identify persistent states in which holding $R$ over interest-bearing cash is poorly compensated. The implementation, however, no longer estimates a tail historical future-return mean. It combines screened continuous components into a direct score and then maps that score into a raw cash weight.

The signed primitive variables are continuous expanding $z$-scores. The cash-rising component is
\begin{equation}
CashRise_t = EZ(\Delta CashYield_{21,t}).
\end{equation}
The compressed-risk-premium block averages five screened low-compensation components:
\begin{align}
LCP_t=\frac{1}{5}\big(&EZ(RS_tLVOL_t)+EZ(RS_tLVOL_tLV_t)+EZ(RS_tLV_t) \nonumber\\
&+EZ(CH_tRS_tLV_t)+EZ(CH_tRS_tLVOL_t)\big),
\end{align}
where $RS$ is risky-sleeve strength, $LVOL$ is low realized risky-sleeve volatility, $LV$ is low VIX, and $CH$ is high cash yield. These products describe states in which the risky sleeve has been strong and calm while cash is becoming more competitive.

The compensation core is
\begin{equation}
CompCore_t=\alpha_{comp}CashRise_t+(1-\alpha_{comp})LCP_t.
\end{equation}
The direct rate headwind is $RH_t=EZ(\Delta TNX_{21,t})$. The rate-path stress block averages four screened interactions:
\begin{equation}
RPS_t=\frac{1}{4}\left(EZ(RH_tVS_t)+EZ(RH_tCRO_t)+EZ(RH_tRU_t)+EZ(DD_tRH_tCRO_t)\right),
\end{equation}
where $VS$ is a VIX spike, $CRO$ is credit risk-off, $RU$ is risky-sleeve underwater, and $DD$ is drawdown depth. The final slow-tail score is
\begin{equation}
SlowScore_t=CompCore_t+\lambda_{rate}RH_t+\lambda_{path}RPS_t,\qquad SlowZ_t=EZ(SlowScore_t).
\end{equation}
The raw cash weight is continuous:
\begin{equation}
CashRaw_t=MaxCash\left[\frac{1}{1+\exp(-SlowZ_t/\tau)}\right]^\gamma .
\end{equation}
If $CashRaw_t \ge 0.30$, entry into cash is immediate. If $CashRaw_t<0.30$, the strategy smoothly returns toward $R$ with an exit smoothing speed of 0.25. The selected full-sample configuration is \code{direct_slow_187}. The grid contains 216 configurations over $\alpha_{comp}$, $\lambda_{rate}$, $\lambda_{path}$, max cash, cash scale, and cash convexity.

\begin{table}[H]
\centering
\caption{Slow-Tail Top Configurations by Selection Score}
\label{tab:slow-top}
\scriptsize
\begin{tabular}{lrrrrrrrr}
\toprule
Config & $\alpha_{comp}$ & $\lambda_{rate}$ & $\lambda_{path}$ & Max Cash & $\tau$ & Power & CAGR & Score\\
\midrule
direct\_slow\_187 & 0.67 & 0.50 & 0.25 & 1.00 & 1.25 & 2 & \pct{17.83} & 0.740\\
direct\_slow\_184 & 0.67 & 0.50 & 0.25 & 1.00 & 1.00 & 2 & \pct{17.60} & 0.733\\
direct\_slow\_214 & 0.67 & 0.50 & 0.50 & 1.00 & 1.25 & 2 & \pct{17.51} & 0.732\\
direct\_slow\_157 & 0.67 & 0.25 & 0.50 & 1.00 & 1.00 & 2 & \pct{17.36} & 0.729\\
direct\_slow\_079 & 0.50 & 0.50 & 0.25 & 1.00 & 1.25 & 2 & \pct{17.35} & 0.724\\
direct\_slow\_127 & 0.67 & 0.25 & 0.25 & 1.00 & 0.75 & 2 & \pct{17.32} & 0.722\\
direct\_slow\_118 & 0.67 & 0.25 & 0.25 & 0.75 & 0.75 & 2 & \pct{17.20} & 0.722\\
direct\_slow\_130 & 0.67 & 0.25 & 0.25 & 1.00 & 1.00 & 2 & \pct{17.25} & 0.721\\
\bottomrule
\end{tabular}
\end{table}

The selected slow-tail structure is not an isolated single parameter point. The leading configurations cluster around $\alpha_{comp}=0.67$, high max cash, low convexity, and positive weights on both rate headwind and rate-path stress (Table~\ref{tab:slow-top}). The implication is that current slow-tail performance is driven by the direct compensation construction rather than by the old analogue-tail parameters.

\begin{table}[H]
\centering
\caption{Slow-Tail Standalone Walk-Forward Validation}
\label{tab:slow-oos-main}
\scriptsize
\begin{tabular}{llrrrrrr}
\toprule
Filter & Mode & CAGR & Max DD & Ann. Turnover & Avg. Cash & 100\% $R$ CAGR & 100\% $R$ Max DD\\
\midrule
Slow-tail & Expanding & \pct{18.79} & \pct{-33.59} & \pct{114.57} & \pct{2.44} & \pct{17.59} & \pct{-33.59}\\
Slow-tail & Rolling & \pct{18.12} & \pct{-33.59} & \pct{138.31} & \pct{2.77} & \pct{17.59} & \pct{-33.59}\\
\bottomrule
\end{tabular}
\end{table}

The first direct walk-forward check supports the slow-tail mechanism but also exposes its boundary. In Table~\ref{tab:slow-oos-main}, expanding and rolling validation both improve CAGR relative to 100\% $R$, yet neither improves the full-window maximum drawdown because the worst drawdown in the main window is a fast crash. This is exactly the boundary of the module: it is a compensation filter for persistent weak $R-C$ regimes, not a crash brake.

\begin{table}[H]
\centering
\caption{Slow-Tail Post-2022 Walk-Forward Validation}
\label{tab:slow-post}
\scriptsize
\begin{tabular}{llrrrrrr}
\toprule
Filter & Mode & CAGR & Max DD & Ann. Turnover & Avg. Cash & 100\% $R$ CAGR & 100\% $R$ Max DD\\
\midrule
Slow-tail & Expanding & \pct{21.19} & \pct{-19.74} & \pct{169.65} & \pct{1.62} & \pct{23.58} & \pct{-19.66}\\
Slow-tail & Rolling & \pct{21.19} & \pct{-19.74} & \pct{169.65} & \pct{1.62} & \pct{23.58} & \pct{-19.66}\\
\bottomrule
\end{tabular}
\end{table}

For slow-tail, the post-2022 test now shows a different trade-off from the previous analogue version. Both expanding and rolling validation hold little average cash and slightly trail 100\% $R$ on CAGR, while drawdown is essentially unchanged (Table~\ref{tab:slow-post}). This weaker post-2022 standalone result is why the current paper treats the redesigned slow-tail branch as a modest 2022 compensation filter rather than as a complete slow-regime solution.

\begin{figure}[H]
\centering
\includegraphics[width=0.98\textwidth]{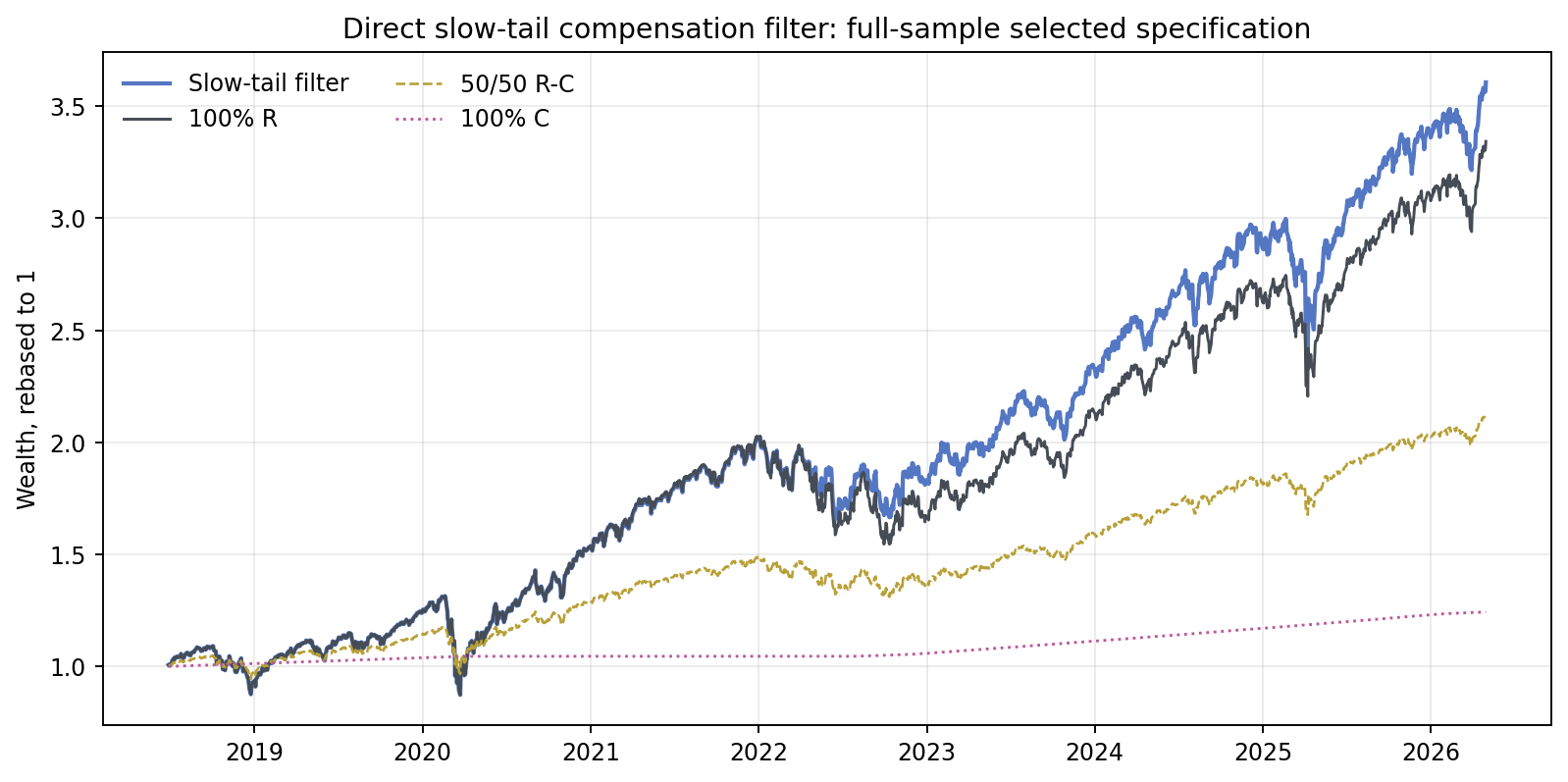}
\caption{Slow-Tail Selected Full-Sample Equity Curves}
\label{fig:slow-tail-equity}
\end{figure}

The equity curve gives the same interpretation as the parameter and OOS tables. The direct slow-tail rule improves long-run wealth modestly and reduces the 2022 drawdown, but the full-sample maximum drawdown remains tied to fast crash dynamics (Figure~\ref{fig:slow-tail-equity}). The module is therefore a slow-compensation filter, not a complete drawdown solution.

\begin{figure}[H]
\centering
\includegraphics[width=0.98\textwidth]{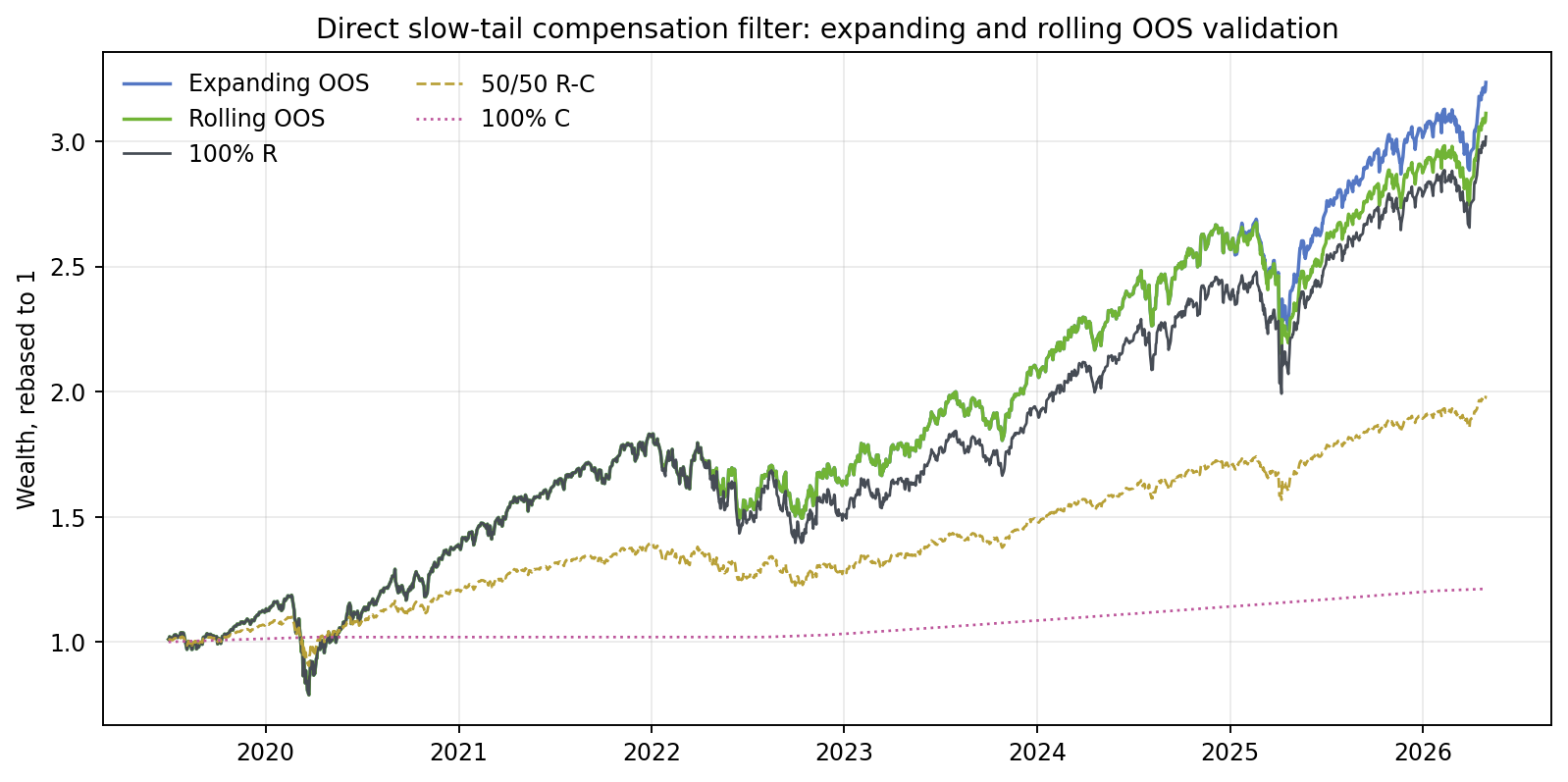}
\caption{Slow-Tail Walk-Forward Equity Curves}
\label{fig:slow-tail-oos}
\end{figure}

The OOS equity paths emphasize timing rather than only terminal wealth. The expanding and rolling slow-tail paths do not defend every drawdown; their value appears when compensation deterioration is persistent enough to be captured by the direct compensation score (Figure~\ref{fig:slow-tail-oos}).

\begin{figure}[H]
\centering
\includegraphics[width=0.98\textwidth]{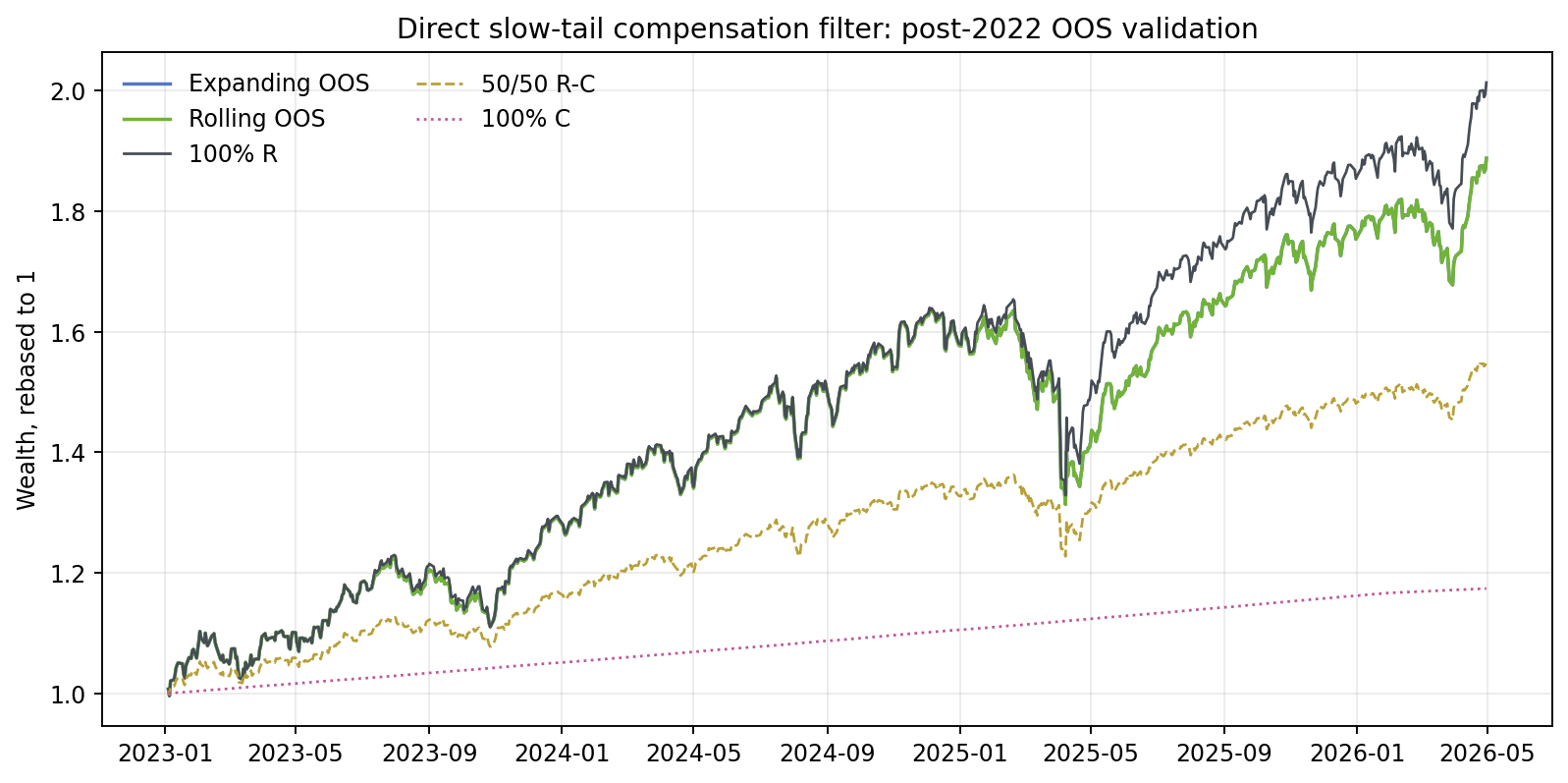}
\caption{Slow-Tail Post-2022 OOS Equity Curves}
\label{fig:slow-tail-post-oos}
\end{figure}

The post-2022 window now shows the cost of making the direct slow-tail branch less aggressive. The strategy does not materially improve the recent drawdown path and trails the strong 100\% $R$ rebound (Figure~\ref{fig:slow-tail-post-oos}). This is one of the reasons the combined policy relies on V-shape for fast stress and does not treat slow-tail as a standalone defense.

\begin{figure}[H]
\centering
\includegraphics[width=0.98\textwidth]{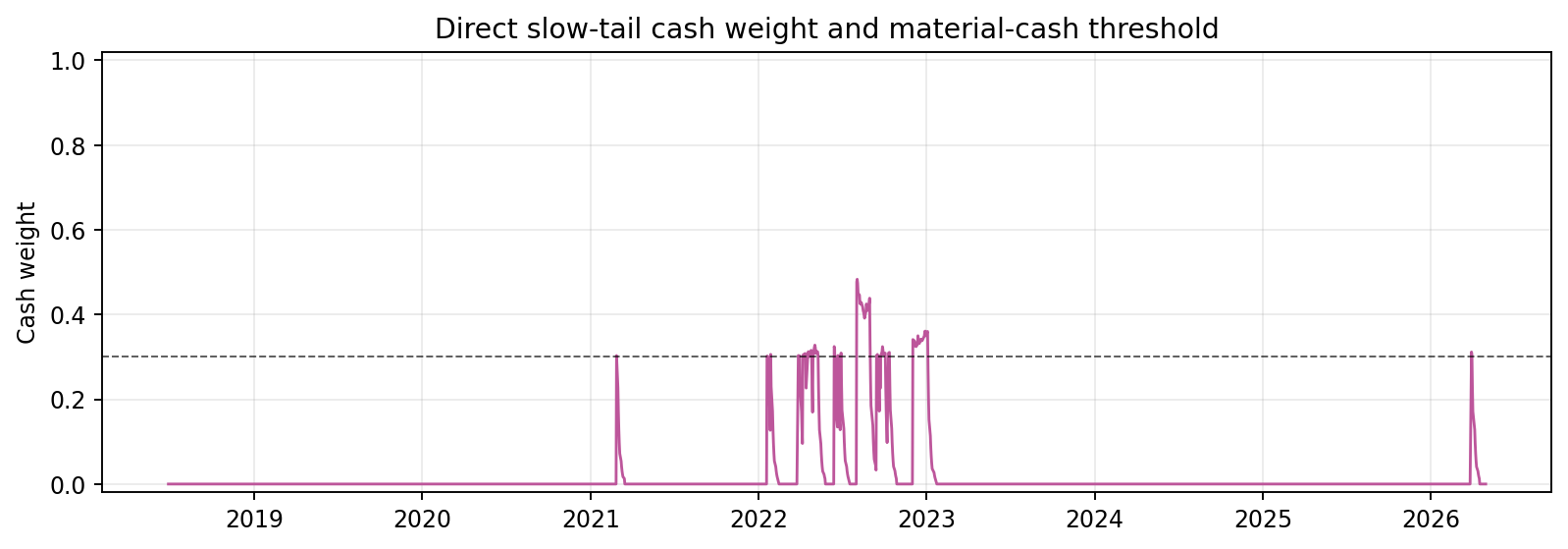}
\caption{Slow-Tail Selected Cash-Weight Path}
\label{fig:slow-tail-cash}
\end{figure}

The selected slow-tail policy is sparse rather than constantly defensive. Cash exposure clusters in slow compensation-deterioration regimes (Figure~\ref{fig:slow-tail-cash}). Entry occurs once the continuous raw cash weight becomes economically material, and the return toward $R$ is gradual because the exit leg is smoothed.

\section{V-Shape Crash-Brake Filter}

\subsection{Policy Construction}

The V-shape filter is designed for fast drawdown events. Unlike slow-tail, it does not estimate a historical analogue mean. It builds a crash-brake score directly from signed market-stress primitives and interaction terms. The signed primitives are
\begin{align}
VL_t &= z_t(VIXPct_{756,t}),&
VS10_t &= z_t(\Delta VIX_{10,t}),\\
RR_t &= z_t(-\Delta TNX_{21,t}),&
RL10_t &= z_t(-RTrailing_{10,t}),\\
CW_t &= z_t(\Delta CS_{21,t}),&
CRO_t &= z_t(-HYGSHYRel_{21,t}),\\
DD_t &= z_t(-RDrawdown_t),&
VR5_t &= z_t(-\Delta VIX_{5,t}),\\
CA5_t &= z_t(HYGTrailing_{5,t}-SHYTrailing_{5,t}).
\end{align}
Here $RR$ is a rate-relief shock, $RL10$ is a 10-day risky-sleeve loss, $CRO$ is credit risk-off, $DD$ is drawdown depth, $VR5$ is VIX relief, and $CA5$ is credit appetite. Candidate products are standardized again before entering the deployable score:
\begin{align}
BrakeVIXLevel_t &= z_t(VL_t),\\
BrakeVIXSpike_t &= z_t(VS10_t),\\
RatePanic_t &= 0.5z_t(RR_tRL10_tVL_t)+0.5z_t(RR_tCRO_t),\\
CreditPanic_t &= z_t(VS10_tCW_t),\\
ReentryOversold_t &= z_t(DD_tVL_t),\\
ReentryRelief_t &= \frac{1}{4}\{z_t(VR5_tCA5_t)+z_t(DD_tVR5_t)\\
&\quad +z_t(VL_tVR5_tCA5_t)+z_t(DD_tVR5_tCA5_t)\}.
\end{align}
The brake score combines a VIX level/spike core with rate-panic and credit-panic interaction terms:
\begin{align}
BrakeScore_t &= VIXCore_t+\lambda_r RatePanic_t+\lambda_c CreditPanic_t,\\
VIXCore_t &= \alpha_{vix}BrakeVIXLevel_t+(1-\alpha_{vix})BrakeVIXSpike_t.
\end{align}
The diagnostic re-entry score is
\begin{equation}
ReentryScore_t=0.8ReentryRelief_t+0.2ReentryOversold_t,
\end{equation}
but it is not used as a separate hard trading gate in the final material policy. It is retained to diagnose whether stress is being replaced by rebound conditions. The brake score is expanding-standardized and mapped into a continuous raw cash weight:
\begin{equation}
CashRaw_t=MaxCash\left[\frac{1}{1+\exp(-BrakeZ_t/\tau)}\right]^\gamma.
\end{equation}
As in the slow-tail policy, cash is activated only when $CashRaw_t\ge 0.30$. Entry into cash is immediate, while exit back to the risky sleeve is smoothed.

The selected full-sample configuration is \pathcode{brake_av0.50_lr0.25_lc0.25_cash0.75_tau1.00_pow5.0_mat0.30_ee1.00_ex0.25}. The grid contains 216 configurations over VIX mix, rate-panic weight, credit-panic weight, max cash, brake scale, and cash convexity.

The V-shape tables in this section are the archived standalone module results. They use the V-shape script's native turnover accounting, which records both the change in risky weight and the offsetting change in cash weight. In the max-cash combination section, the same selected V-shape weight path is recomputed on the shared daily panel using the final one-way $R$-weight turnover convention. That recomputation is the correct comparison for the combined portfolio.

\begin{table}[H]
\centering
\caption{V-Shape Top Configurations by Selection Score}
\label{tab:v-top}
\scriptsize
\begin{tabular}{rrrrrrrr}
\toprule
$\alpha_{vix}$ & $\lambda_r$ & $\lambda_c$ & Max Cash & $\tau$ & Power & CAGR & Score\\
\midrule
0.50 & 0.25 & 0.25 & 0.75 & 1.00 & 5 & \pct{17.32} & 0.769\\
0.50 & 0.50 & 0.25 & 1.00 & 1.00 & 5 & \pct{17.07} & 0.763\\
0.50 & 0.50 & 0.25 & 0.75 & 1.00 & 5 & \pct{17.44} & 0.758\\
0.67 & 0.25 & 0.25 & 0.75 & 1.00 & 5 & \pct{17.64} & 0.756\\
0.50 & 0.25 & 0.25 & 0.50 & 1.00 & 4 & \pct{17.67} & 0.754\\
0.50 & 0.25 & 0.25 & 0.50 & 0.75 & 5 & \pct{17.29} & 0.752\\
0.67 & 0.50 & 0.25 & 0.75 & 1.00 & 5 & \pct{17.57} & 0.744\\
0.50 & 0.25 & 0.25 & 0.50 & 1.00 & 3 & \pct{17.27} & 0.743\\
\bottomrule
\end{tabular}
\end{table}

The selected V-shape rules have a clear crash-brake shape. In Table~\ref{tab:v-top}, the leading configurations prefer a balanced VIX level/spike core and modest rate/credit panic weights. They also favor high cash convexity, which means cash exposure is not increased linearly for ordinary stress. Instead, the rule waits for an extreme brake score before allocating substantial cash. This is consistent with the intended use case: the module should be quiet most of the time and aggressive only when crash-like conditions appear.

\begin{table}[H]
\centering
\caption{V-Shape Event Diagnostics}
\label{tab:v-events}
\scriptsize
\begin{tabular}{lrrrrrr}
\toprule
Event & Strategy Ret. & 100\% $R$ Ret. & Excess & Avg. Cash & Max Cash & Days\\
\midrule
COVID crash & \pct{-14.66} & \pct{-33.17} & \pct{18.50} & \pct{58.64} & \pct{74.88} & 24\\
April 2025 crash & \pct{-10.17} & \pct{-14.40} & \pct{4.23} & \pct{19.33} & \pct{74.32} & 11\\
Slow 2022 window & \pct{-12.43} & \pct{-12.43} & \pct{0.00} & \pct{0.00} & \pct{0.00} & 22\\
COVID recovery & \pct{19.75} & \pct{27.21} & \pct{-7.46} & \pct{12.50} & \pct{67.54} & 28\\
April 2025 recovery & \pct{5.49} & \pct{9.90} & \pct{-4.41} & \pct{41.91} & \pct{74.54} & 16\\
\bottomrule
\end{tabular}
\end{table}

Event behavior gives the clearest description of the V-shape module. During the COVID crash, the strategy cuts the drawdown-period loss from -33.17\% to -14.66\%, with average cash exposure of 58.64\% (Table~\ref{tab:v-events}). During the April 2025 crash, it also reduces loss, though less dramatically, because the event is shorter and the rebound is fast. The negative recovery rows are the price of this protection: once the rule holds cash during a V-shaped rebound, it can lag the risky sleeve. The zero cash exposure in the slow 2022 window confirms that this module does not solve the 2022-style problem.

\begin{table}[H]
\centering
\caption{V-Shape Standalone Walk-Forward Validation}
\label{tab:v-oos-main}
\scriptsize
\begin{tabular}{llrrrrrr}
\toprule
Filter & Mode & CAGR & Max DD & Ann. Turnover & Avg. Cash & 100\% $R$ CAGR & 100\% $R$ Max DD\\
\midrule
V-shape & Expanding & \pct{15.59} & \pct{-25.98} & \pct{513.18} & \pct{3.92} & \pct{16.09} & \pct{-33.59}\\
V-shape & Rolling & \pct{14.88} & \pct{-26.37} & \pct{585.82} & \pct{4.65} & \pct{16.09} & \pct{-33.59}\\
\bottomrule
\end{tabular}
\end{table}

The standalone OOS evidence should therefore be read through a drawdown lens, not as an alpha table. Table~\ref{tab:v-oos-main} shows that both expanding and rolling versions reduce maximum drawdown by roughly seven percentage points relative to 100\% $R$, but both trail the risky sleeve on CAGR. That trade-off is acceptable only if the module is combined with a separate return-preserving or slow-tail component, which is exactly why the max-cash combination is tested later.

\begin{table}[H]
\centering
\caption{V-Shape Post-2022 Walk-Forward Validation}
\label{tab:v-post}
\scriptsize
\begin{tabular}{llrrrrrr}
\toprule
Filter & Mode & CAGR & Max DD & Ann. Turnover & Avg. Cash & 100\% $R$ CAGR & 100\% $R$ Max DD\\
\midrule
V-shape & Expanding & \pct{10.32} & \pct{-25.97} & \pct{521.66} & \pct{3.15} & \pct{12.47} & \pct{-23.78}\\
V-shape & Rolling & \pct{8.35} & \pct{-26.36} & \pct{661.05} & \pct{4.67} & \pct{12.47} & \pct{-23.78}\\
\bottomrule
\end{tabular}
\end{table}

The post-2022 V-shape result is intentionally weaker. The period contains more slow deterioration than sudden crash onset, so the V-shape brake tends either to stay inactive or to enter too late relative to a slow drawdown (Table~\ref{tab:v-post}). This is not a failure of the module's stated purpose; it is evidence that the module should not be used as the only cash filter.

\begin{figure}[H]
\centering
\includegraphics[width=0.98\textwidth]{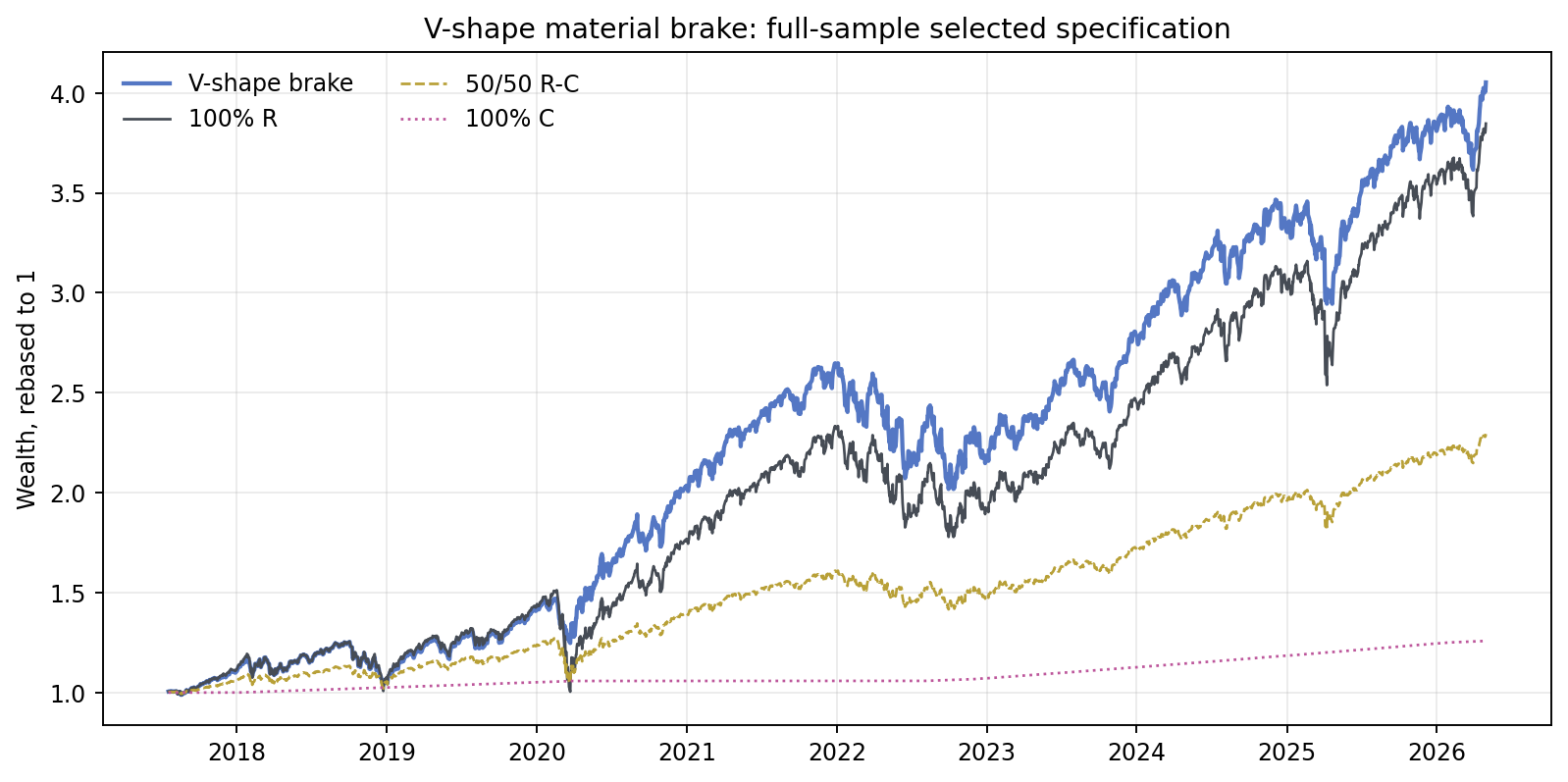}
\caption{V-Shape Selected Full-Sample Equity Curves}
\label{fig:v-main}
\end{figure}

The V-shape module's main trade-off is visible in the selected full-sample path. The strategy lowers fast-crash drawdowns with low average cash exposure, but it can lag during very fast recoveries because cash must be released after stress has already started to normalize (Figure~\ref{fig:v-main}). This explains why the module is evaluated as a crash brake rather than as a standalone return enhancer.

\begin{figure}[H]
\centering
\includegraphics[width=0.98\textwidth]{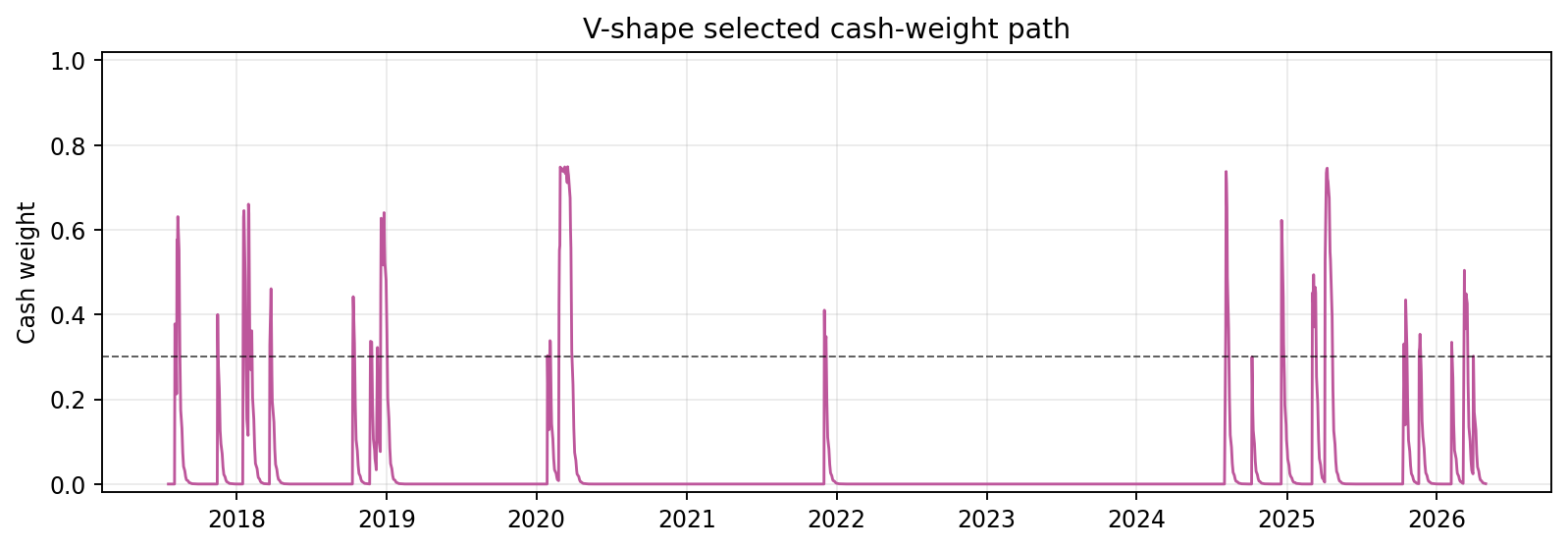}
\caption{V-Shape Selected Cash-Weight Path}
\label{fig:v-cash-weight}
\end{figure}

The selected cash-weight path shows how sparse the V-shape brake is in normal markets. Cash exposure remains close to zero most of the time and rises only around short stress episodes, which is consistent with the module's intended role as a temporary crash brake rather than a slow-cycle defensive allocator (Figure~\ref{fig:v-cash-weight}).

\begin{figure}[H]
\centering
\includegraphics[width=0.98\textwidth]{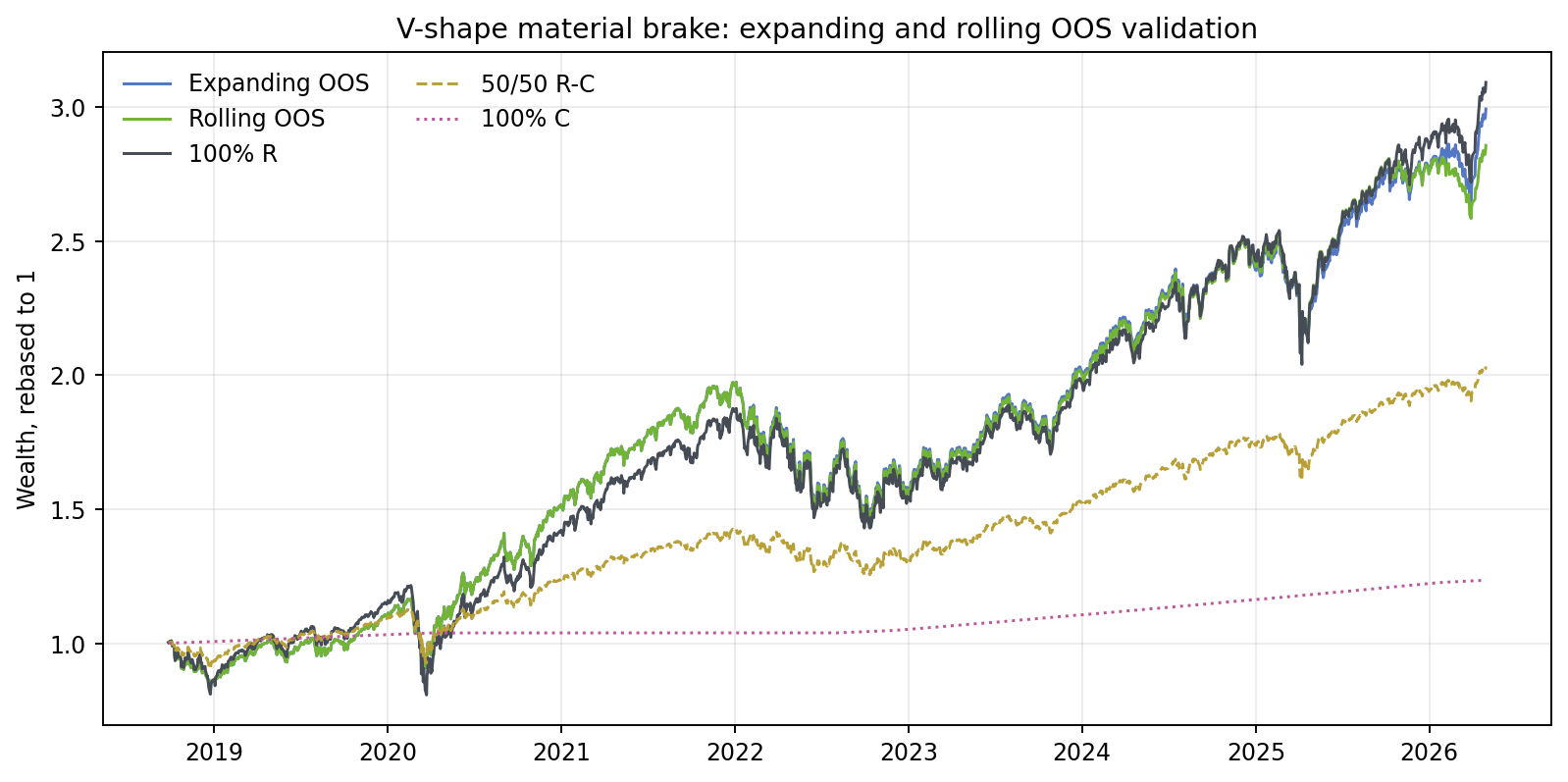}
\caption{V-Shape Walk-Forward Equity Curves}
\label{fig:v-oos}
\end{figure}

The OOS paths reinforce the same trade-off. The V-shape strategies reduce drawdown relative to the risky sleeve, but they do not consistently outperform on terminal wealth (Figure~\ref{fig:v-oos}). This is why the paper later combines V-shape with slow-tail rather than presenting it as a complete cash-allocation policy.

\begin{figure}[H]
\centering
\includegraphics[width=0.98\textwidth]{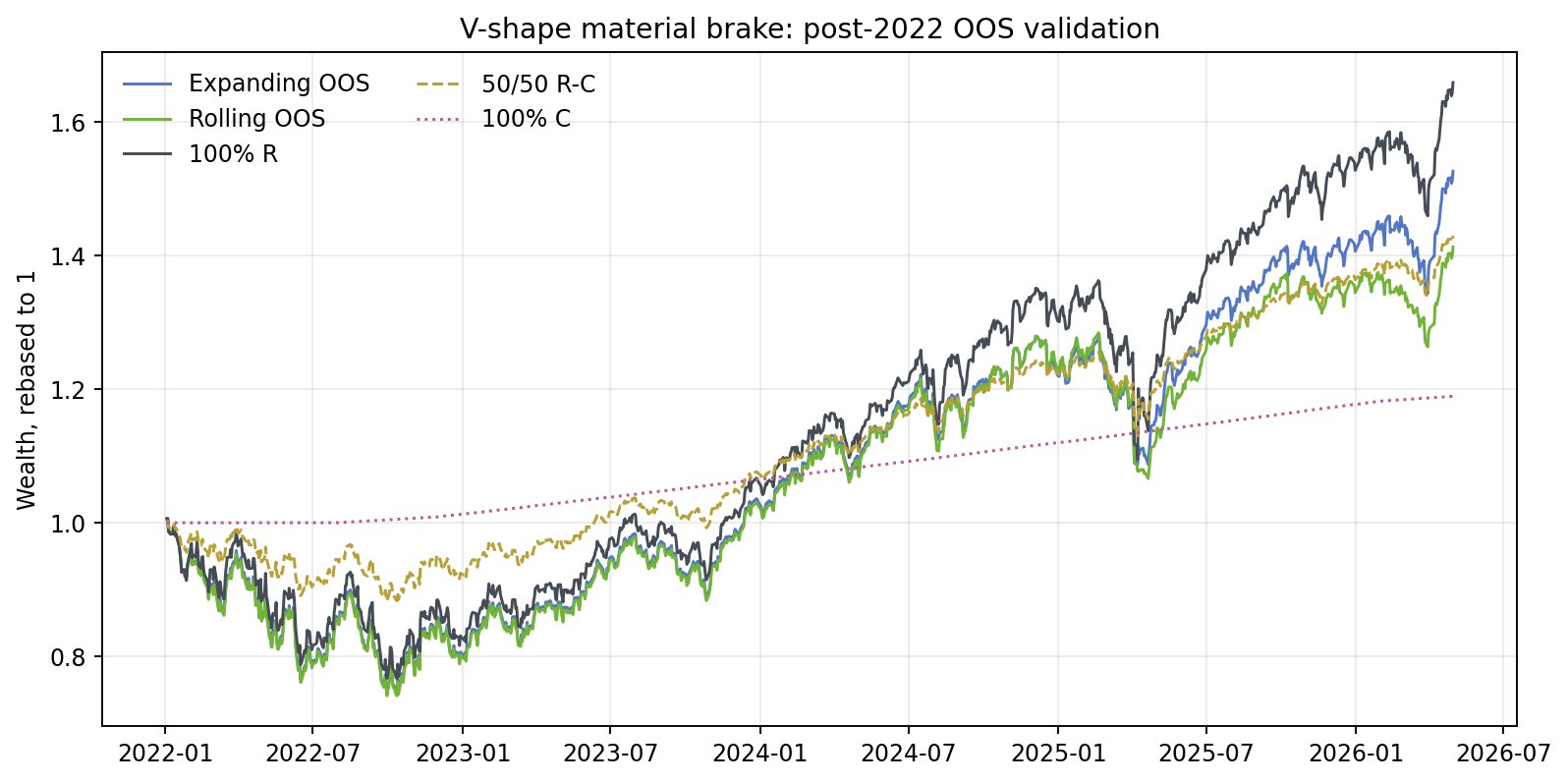}
\caption{V-Shape Post-2022 OOS Equity Curves}
\label{fig:v-post-oos}
\end{figure}

The post-2022 path confirms the limits of the crash-brake design. The window is not one clean V-shaped crash, so the V-shape module can either stay inactive during slow deterioration or create rebound opportunity cost after stress spikes (Figure~\ref{fig:v-post-oos}). This visual result is consistent with the event diagnostics in Table~\ref{tab:v-events}: the module is useful for fast stress, but it is not a complete cash-allocation policy by itself.

\section{Max-Cash Combination}

The two filters are combined by a fixed conservative rule:
\begin{equation}
w^C_t=\max(w^{C,slow}_t,w^{C,vshape}_t),\qquad w^R_t=1-w^C_t.
\end{equation}
This rule does not select between models. It treats the two filters as independent risk alarms. If either module demands a larger cash allocation, the final portfolio uses that larger cash weight.

\begin{table}[H]
\centering
\caption{Max-Cash Selected-Weight Combination, Common Window}
\label{tab:combo-full}
\scriptsize
\begin{tabular}{lrrrrrr}
\toprule
Method & CAGR & Max DD & Ann. Turnover & Avg. Cash & Max Cash & Final Wealth\\
\midrule
Max-cash & \pct{18.83} & \pct{-18.05} & \pct{355.82} & \pct{5.56} & \pct{74.88} & 4.54\\
Slow-tail only & \pct{17.63} & \pct{-33.59} & \pct{96.68} & \pct{1.96} & \pct{48.33} & 4.15\\
V-shape only & \pct{17.63} & \pct{-23.77} & \pct{266.20} & \pct{3.65} & \pct{74.88} & 4.15\\
100\% $R$ & \pct{16.62} & \pct{-33.59} & \pct{0.00} & \pct{0.00} & \pct{0.00} & 3.85\\
\bottomrule
\end{tabular}
\end{table}

The combination is economically different from either component alone. Table~\ref{tab:combo-full} recomputes slow-tail, V-shape, max-cash, and the benchmarks on the same common daily panel with the final one-way turnover convention. This is why the V-shape-only CAGR and turnover differ from the standalone V-shape table above even though the selected weight path is the same. Under this common convention, the redesigned slow-tail module improves return but does not reduce the main maximum drawdown; V-shape reduces drawdown but is not a 2022 slow-tail solution. The max-cash rule keeps the stronger defensive action each day, so its maximum drawdown is lower than either component and its CAGR remains above 100\% $R$. The stricter OOS-weight combination is reported next.

\begin{table}[H]
\centering
\caption{Key Event Diagnostics for Max-Cash Combination}
\label{tab:combo-events}
\scriptsize
\begin{tabular}{lrrrrr}
\toprule
Event & Policy Ret. & 100\% $R$ Ret. & Policy DD & 100\% $R$ DD & Avg. Cash\\
\midrule
COVID crash, 2020-02-19--2020-03-23 & \pct{-14.58} & \pct{-33.17} & \pct{-15.11} & \pct{-33.59} & \pct{58.64}\\
Calendar 2022 & \pct{-8.95} & \pct{-17.06} & \pct{-18.05} & \pct{-23.78} & \pct{15.60}\\
Slow 2022, 2022-08-25--2022-09-26 & \pct{-8.78} & \pct{-12.43} & \pct{-9.59} & \pct{-13.75} & \pct{24.38}\\
April 2025 crash, 2025-03-25--2025-04-08 & \pct{-10.10} & \pct{-14.40} & \pct{-10.11} & \pct{-14.41} & \pct{19.33}\\
April 2025 crash plus recovery & \pct{-4.66} & \pct{-4.20} & \pct{-10.11} & \pct{-14.41} & \pct{31.11}\\
\bottomrule
\end{tabular}
\end{table}

The event decomposition clarifies why a simple max-cash rule is enough. In Table~\ref{tab:combo-events}, the COVID crash benefit comes almost entirely from the fixed V-shape module, while the 2022 benefit comes from the redesigned slow-tail branch. The current direct slow-tail version no longer eliminates the 2022 drawdown as aggressively as the previous analogue version, but it still improves both the full calendar 2022 and the August--September slow-tail window. The combined rule does not need to identify the event type in advance; it simply takes the larger cash weight from whichever module is active.

\begin{table}[H]
\centering
\caption{Max-Cash Rule Applied to Sub-Strategy OOS Weights}
\label{tab:combo-oos}
\scriptsize
\begin{tabular}{lllrrrr}
\toprule
OOS Window & Mode & Method & CAGR & Max DD & Ann. Turnover & Avg. Cash\\
\midrule
main-oos & Expanding & Max-cash & \pct{19.35} & \pct{-22.05} & \pct{305.09} & \pct{5.22}\\
main-oos & Expanding & 100\% $R$ & \pct{17.59} & \pct{-33.59} & \pct{0.00} & \pct{0.00}\\
main-oos & Rolling & Max-cash & \pct{18.50} & \pct{-22.05} & \pct{351.81} & \pct{6.15}\\
main-oos & Rolling & 100\% $R$ & \pct{17.59} & \pct{-33.59} & \pct{0.00} & \pct{0.00}\\
post2022-oos & Expanding & Max-cash & \pct{20.76} & \pct{-15.53} & \pct{397.21} & \pct{4.56}\\
post2022-oos & Expanding & 100\% $R$ & \pct{23.58} & \pct{-19.66} & \pct{0.00} & \pct{0.00}\\
post2022-oos & Rolling & Max-cash & \pct{18.91} & \pct{-17.26} & \pct{470.06} & \pct{6.24}\\
post2022-oos & Rolling & 100\% $R$ & \pct{23.58} & \pct{-19.66} & \pct{0.00} & \pct{0.00}\\
\bottomrule
\end{tabular}
\end{table}

The OOS-weight combination is the more conservative comparison. Here the component weights are not the selected full-sample paths; they are generated by each component's own expanding or rolling walk-forward selection (Table~\ref{tab:combo-oos}). The main OOS window improves both CAGR and drawdown for both expanding and rolling versions. In the post-2022 window, both expanding and rolling lower drawdown but sacrifice CAGR relative to 100\% $R$. This is the key cost of the cash overlay during a strong risky-sleeve rebound.

\begin{table}[H]
\centering
\caption{Component Trigger Overlap, 1\% Cash-Weight Threshold}
\label{tab:overlap}
\scriptsize
\begin{tabular}{lrrr}
\toprule
Bucket & Days & Sample Share & Avg. Combined Cash\\
\midrule
Slow-tail only & 189 & \pct{8.56} & \pct{22.07}\\
V-shape only & 347 & \pct{15.72} & \pct{22.74}\\
Both active & 13 & \pct{0.59} & \pct{11.46}\\
Combined active & 549 & \pct{24.86} & \pct{22.24}\\
\bottomrule
\end{tabular}
\end{table}

Redundancy is low despite the simple combination rule. At the reported active-state threshold, Table~\ref{tab:overlap} shows that both modules are active together on only 13 trading days, or 0.59\% of the common window. The combined active share of 24.86\% is therefore not produced by two copies of the same signal. The modules fire in different market states, which supports the max-cash combination rule.

\begin{figure}[H]
\centering
\includegraphics[width=0.98\textwidth]{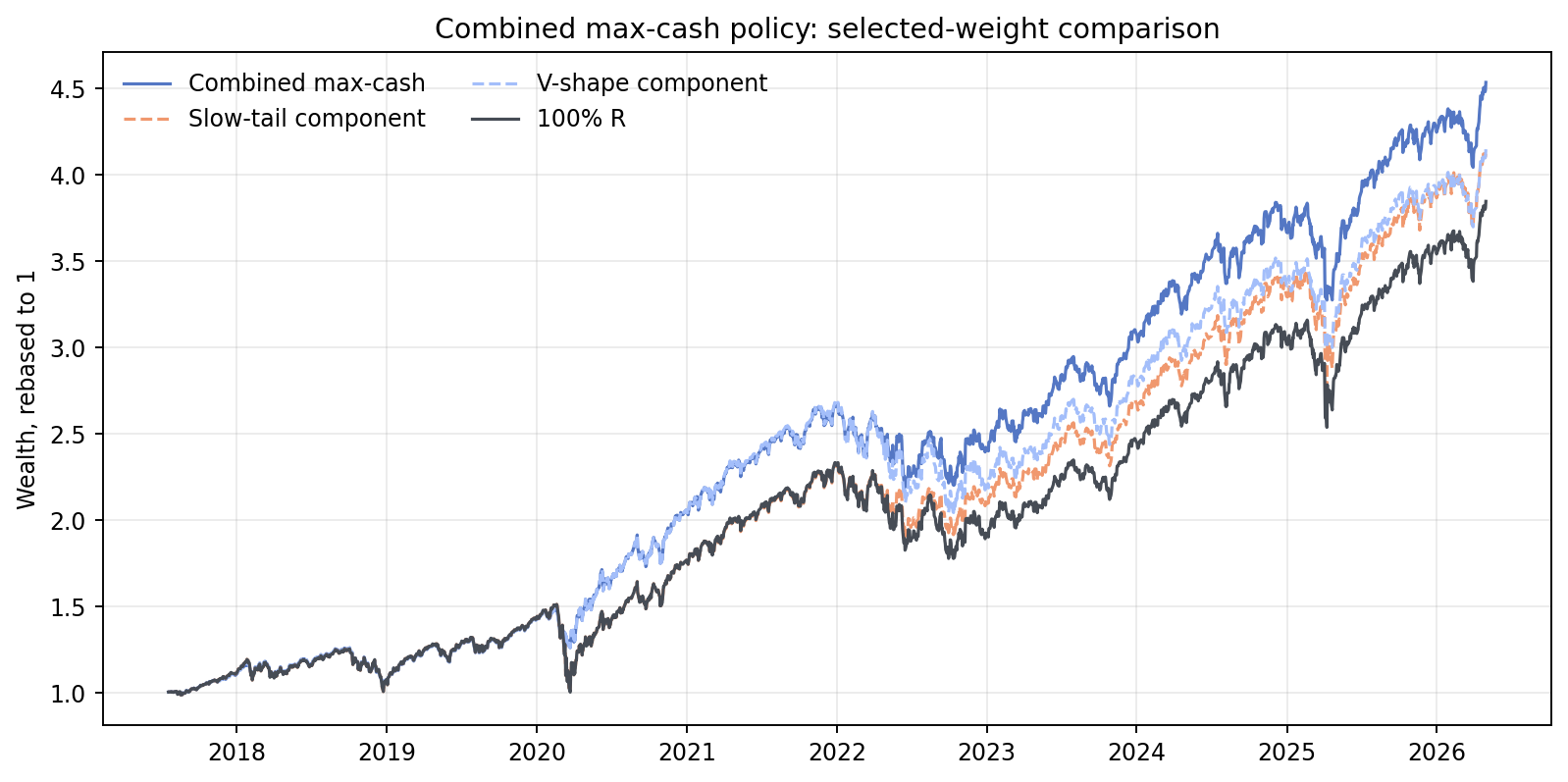}
\caption{Selected-Weight Max-Cash Combination Equity Curves}
\label{fig:combo-full}
\end{figure}

The selected-weight equity paths show how the upper-envelope rule works in practice. The combined line is not an average of the two filters; it follows the larger cash demand each day (Figure~\ref{fig:combo-full}). This is why the combined path can keep the slow-tail 2022 benefit and the V-shape fast-crash benefit in the same portfolio.

\begin{figure}[H]
\centering
\includegraphics[width=0.98\textwidth]{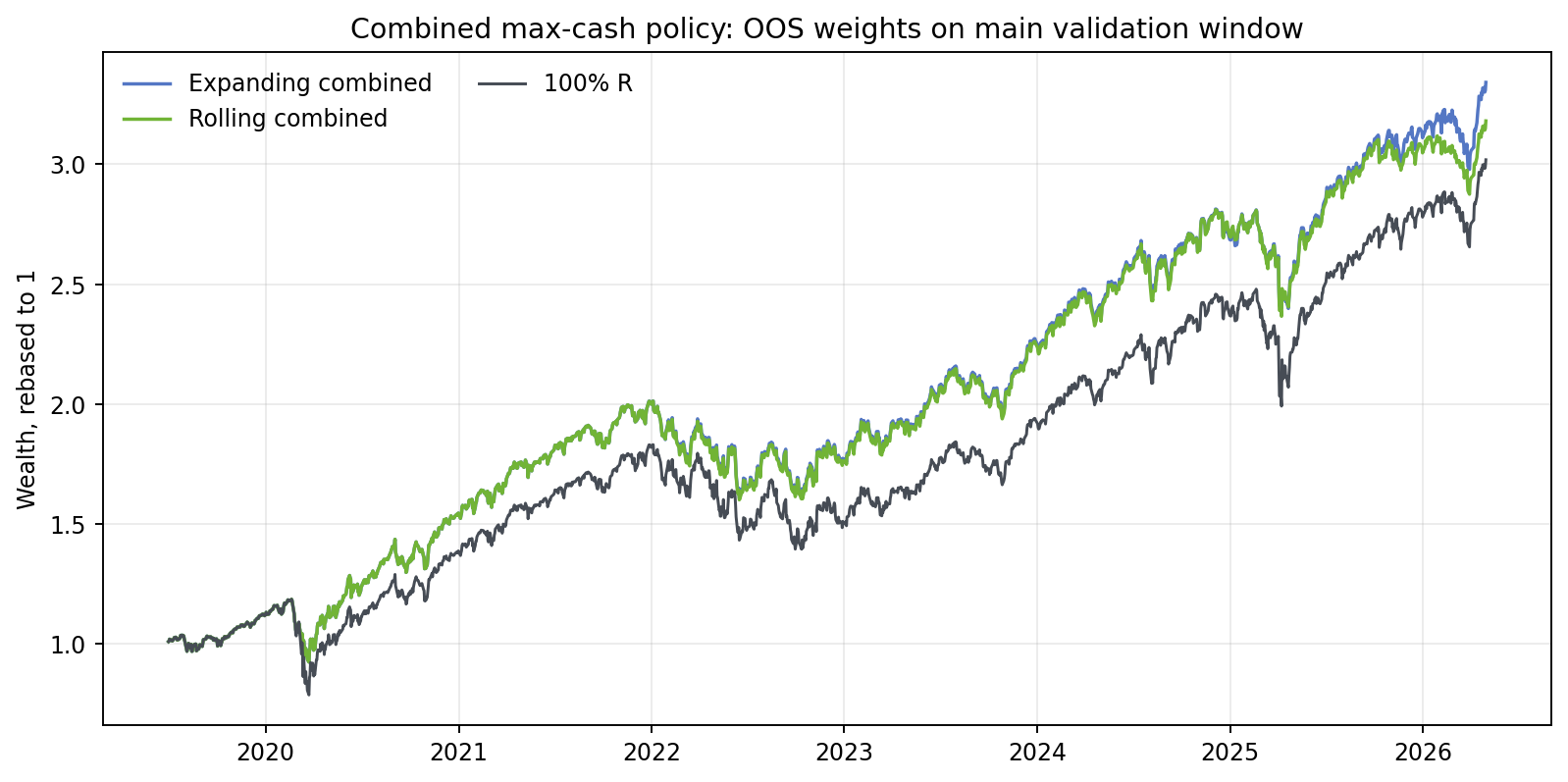}
\caption{Main OOS Max-Cash Combination Equity Curves}
\label{fig:combo-oos}
\end{figure}

The stricter OOS-weight construction is shown separately. Each component first selects parameters using only past data, and the max-cash rule is then applied to those OOS weights (Figure~\ref{fig:combo-oos}). This checks whether the modular combination survives after removing the selected full-sample component paths.

\begin{figure}[H]
\centering
\includegraphics[width=0.98\textwidth]{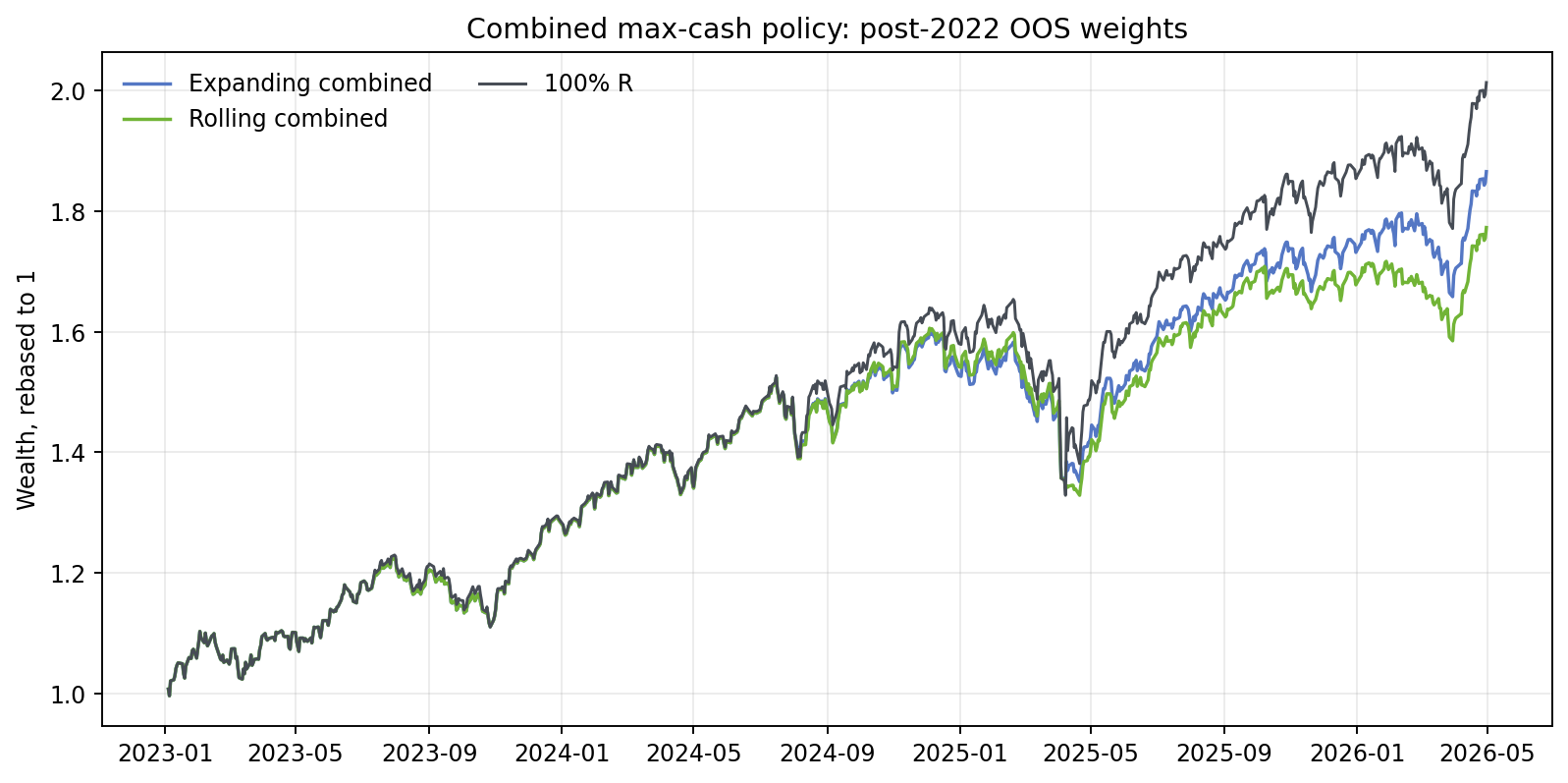}
\caption{Post-2022 OOS Max-Cash Combination Equity Curves}
\label{fig:combo-post-oos}
\end{figure}

The post-2022 OOS-weight comparison separates drawdown control from return capture. Both expanding and rolling versions reduce drawdown versus 100\% $R$, but both trail the strong risky-sleeve rebound in terminal wealth (Figure~\ref{fig:combo-post-oos}). This visual distinction matches the numerical comparison in Table~\ref{tab:combo-oos}.

\begin{figure}[H]
\centering
\includegraphics[width=0.98\textwidth]{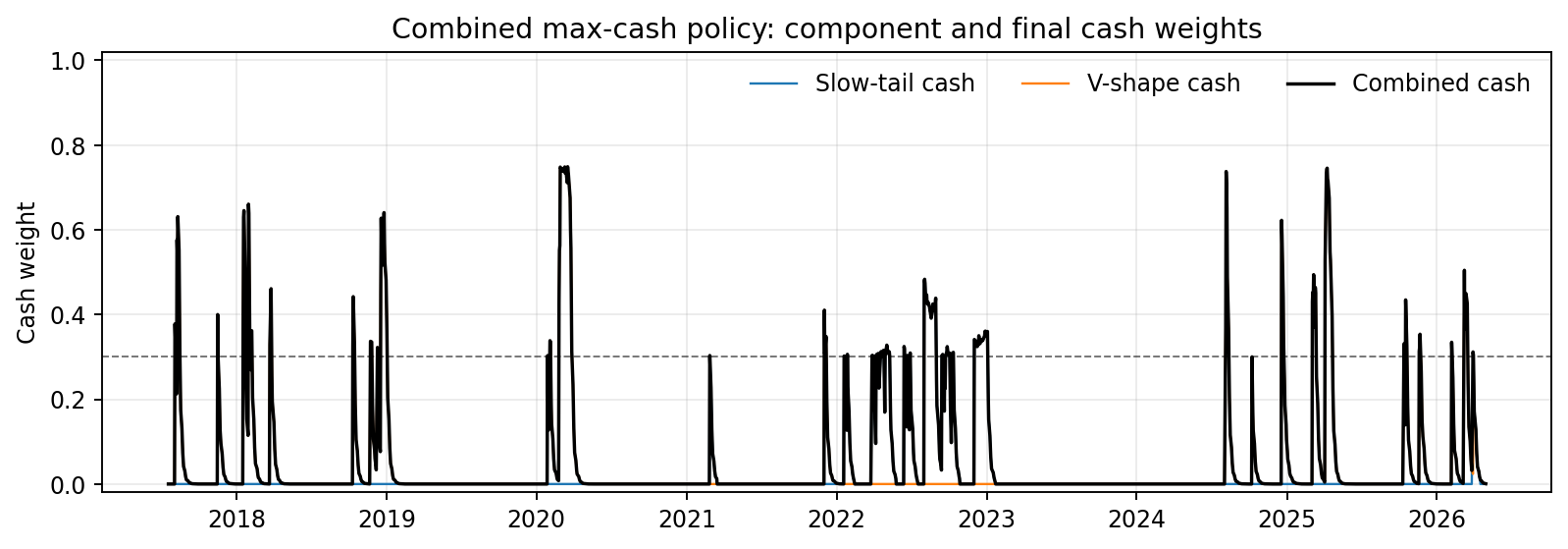}
\caption{Selected-Weight Component and Max-Cash Allocations}
\label{fig:combo-cash}
\end{figure}

The cash-weight path makes the mechanics behind the combined equity curve explicit. The combined cash weight is the upper envelope of the slow-tail and V-shape cash weights (Figure~\ref{fig:combo-cash}). This preserves specialized alarms: slow-tail can dominate during persistent weak compensation, while V-shape can dominate during fast stress events.

\section{Rolling Start-Date Sensitivity}

The companion style-timing study reports start-date sensitivity because a single OOS start can hide dependence on block boundaries \cite{xiong2026continuous}. The same diagnostic is repeated here. For each requested start date, the two component policies reselect the full score and position-parameter vector in each 63-trading-day OOS block. The rolling version uses the most recent 756 trading days as the training window. The max-cash combination is then rebuilt from those two component OOS weight paths.

The rolling result is economically useful but not uniform. Because the direct slow-tail module requires a 252-trading-day warmup inside each requested-start experiment, the first effective OOS date is about one year after the requested start for the combined path. For requested starts through 2019, rolling max-cash improves both CAGR and maximum drawdown relative to 100\% $R$. For the 2020 and 2022 requested starts, it still reduces drawdown but gives up CAGR, which is consistent with a short training window that becomes more defensive after recent stress.

\begin{table}[H]
\centering
\caption{Rolling Start-Date Sensitivity: Start-to-End CAGR and Drawdown}
\label{tab:rolling-start-sens}
\begingroup
\tiny
\setlength{\tabcolsep}{3pt}
\resizebox{\textwidth}{!}{%
\begin{tabular}{lrrrrrrr}
\toprule
Requested Start & Max-Cash CAGR & Sharpe & Max DD & CAGR Diff & MDD Improvement & 100\% $R$ CAGR & 100\% $R$ Max DD\\
\midrule
2018-06-28 & \pct{18.50} & 1.09 & \pct{-22.05} & \pct{0.90} & \pct{11.53} & \pct{17.59} & \pct{-33.59}\\
2018-07-30 & \pct{18.61} & 1.09 & \pct{-22.05} & \pct{1.35} & \pct{11.53} & \pct{17.27} & \pct{-33.59}\\
2018-08-29 & \pct{19.50} & 1.14 & \pct{-22.05} & \pct{1.37} & \pct{11.53} & \pct{18.13} & \pct{-33.59}\\
2018-10-29 & \pct{19.36} & 1.12 & \pct{-22.05} & \pct{1.63} & \pct{11.53} & \pct{17.74} & \pct{-33.59}\\
2019-01-02 & \pct{17.62} & 1.04 & \pct{-22.05} & \pct{0.63} & \pct{11.53} & \pct{16.99} & \pct{-33.59}\\
2020-01-02 & \pct{15.20} & 1.01 & \pct{-19.18} & \pct{-0.64} & \pct{4.60} & \pct{15.84} & \pct{-23.78}\\
2022-01-03 & \pct{18.91} & 1.40 & \pct{-17.26} & \pct{-4.67} & \pct{2.41} & \pct{23.58} & \pct{-19.66}\\
\bottomrule
\end{tabular}
}
\endgroup
\end{table}

Across requested starts, the rolling version is robust for drawdown control but not for return dominance. Table~\ref{tab:rolling-start-sens} shows positive CAGR improvement for 2018--2019 requested starts, while the 2020 and 2022 requested starts lag 100\% $R$ after still reducing drawdown. This pattern means the rolling window is useful as a recent-regime adapter, but it can become too anchored to recent stress or miss part of a strong rebound.

\begin{table}[H]
\centering
\caption{Rolling Common-Window Sensitivity, 2022-01-03 to 2026-04-30}
\label{tab:rolling-common-sens}
\begingroup
\tiny
\setlength{\tabcolsep}{3pt}
\resizebox{\textwidth}{!}{%
\begin{tabular}{lrrrrrrr}
\toprule
Requested Start & Max-Cash CAGR & Max DD & 100\% $R$ CAGR & 50/50 $R-C$ CAGR & 100\% $C$ CAGR & CAGR Diff & MDD Improvement\\
\midrule
2018-06-28 & \pct{11.38} & \pct{-20.42} & \pct{12.47} & \pct{8.65} & \pct{4.11} & \pct{-1.10} & \pct{3.36}\\
2018-07-30 & \pct{12.34} & \pct{-20.83} & \pct{12.47} & \pct{8.65} & \pct{4.11} & \pct{-0.14} & \pct{2.95}\\
2018-08-29 & \pct{12.35} & \pct{-19.70} & \pct{12.47} & \pct{8.65} & \pct{4.11} & \pct{-0.13} & \pct{4.08}\\
2018-10-29 & \pct{13.23} & \pct{-18.33} & \pct{12.47} & \pct{8.65} & \pct{4.11} & \pct{0.76} & \pct{5.45}\\
2019-01-02 & \pct{12.04} & \pct{-19.18} & \pct{12.47} & \pct{8.65} & \pct{4.11} & \pct{-0.43} & \pct{4.60}\\
2020-01-02 & \pct{12.05} & \pct{-19.18} & \pct{12.47} & \pct{8.65} & \pct{4.11} & \pct{-0.43} & \pct{4.60}\\
2022-01-03 & \pct{18.91} & \pct{-17.26} & \pct{23.58} & \pct{14.23} & \pct{4.97} & \pct{-4.67} & \pct{2.41}\\
\bottomrule
\end{tabular}
}
\endgroup
\end{table}

Putting all rolling variants on the same post-2022 evaluation window isolates the effect of the requested start from the effect of different calendar samples. The common-window comparison in Table~\ref{tab:rolling-common-sens} shows the cost of the current direct slow-tail design most clearly: drawdown improvement is generally positive, but CAGR superiority appears only for one requested-start variant and is lost when the risky sleeve rebounds strongly.

\begin{figure}[H]
\centering
\includegraphics[width=0.98\textwidth]{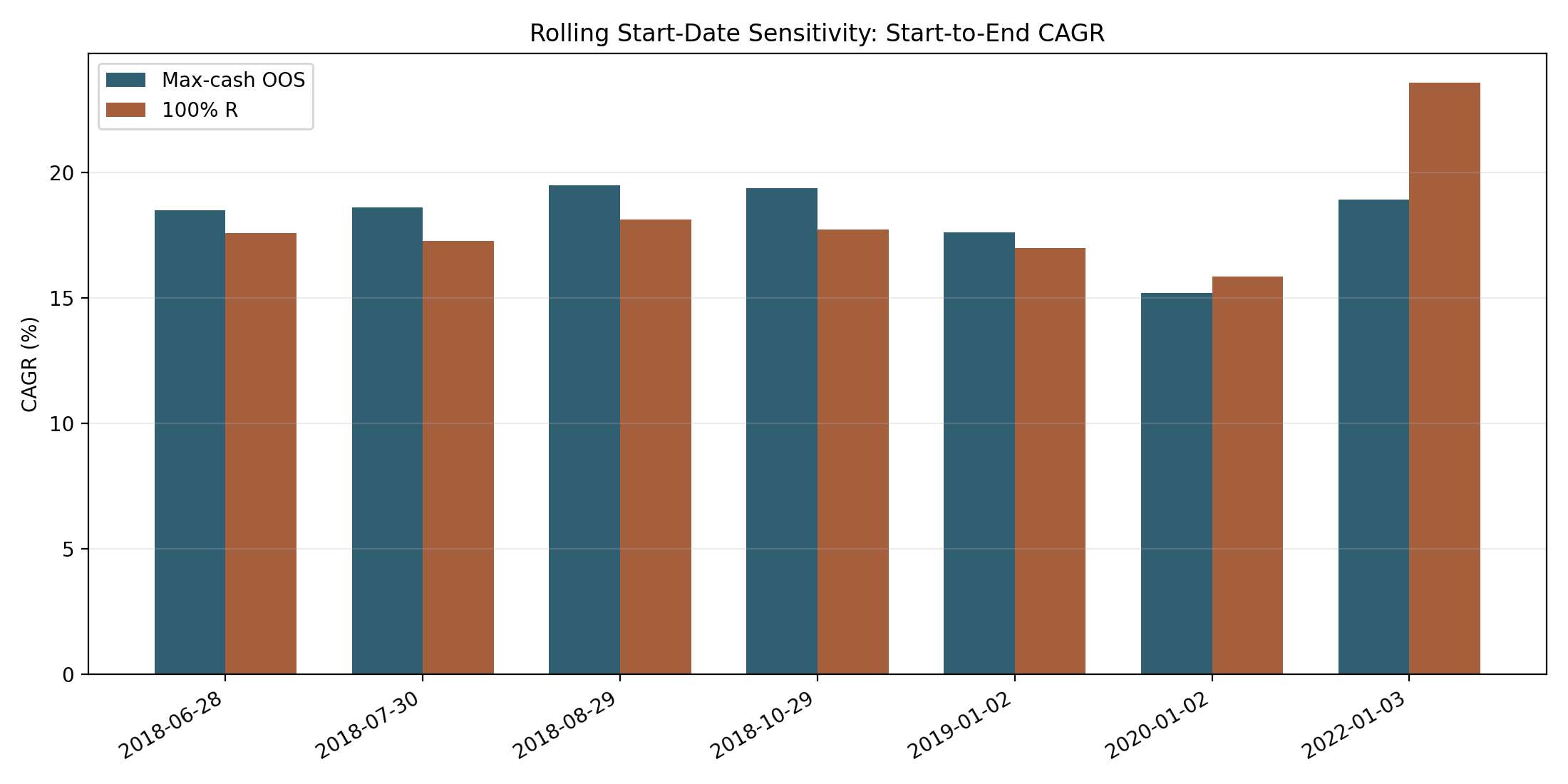}
\caption{Rolling Start-Date Sensitivity: Start-to-End CAGR}
\label{fig:rolling-start-bar}
\end{figure}

The rolling start-to-end CAGR comparison is visually asymmetric. The rolling max-cash rule beats 100\% $R$ for requested starts through 2019, while later requested starts are weaker (Figure~\ref{fig:rolling-start-bar}). This highlights the same caution as Table~\ref{tab:rolling-start-sens}: rolling selection adapts quickly but may become over-defensive after a stress-heavy training window.

\begin{figure}[H]
\centering
\includegraphics[width=0.98\textwidth]{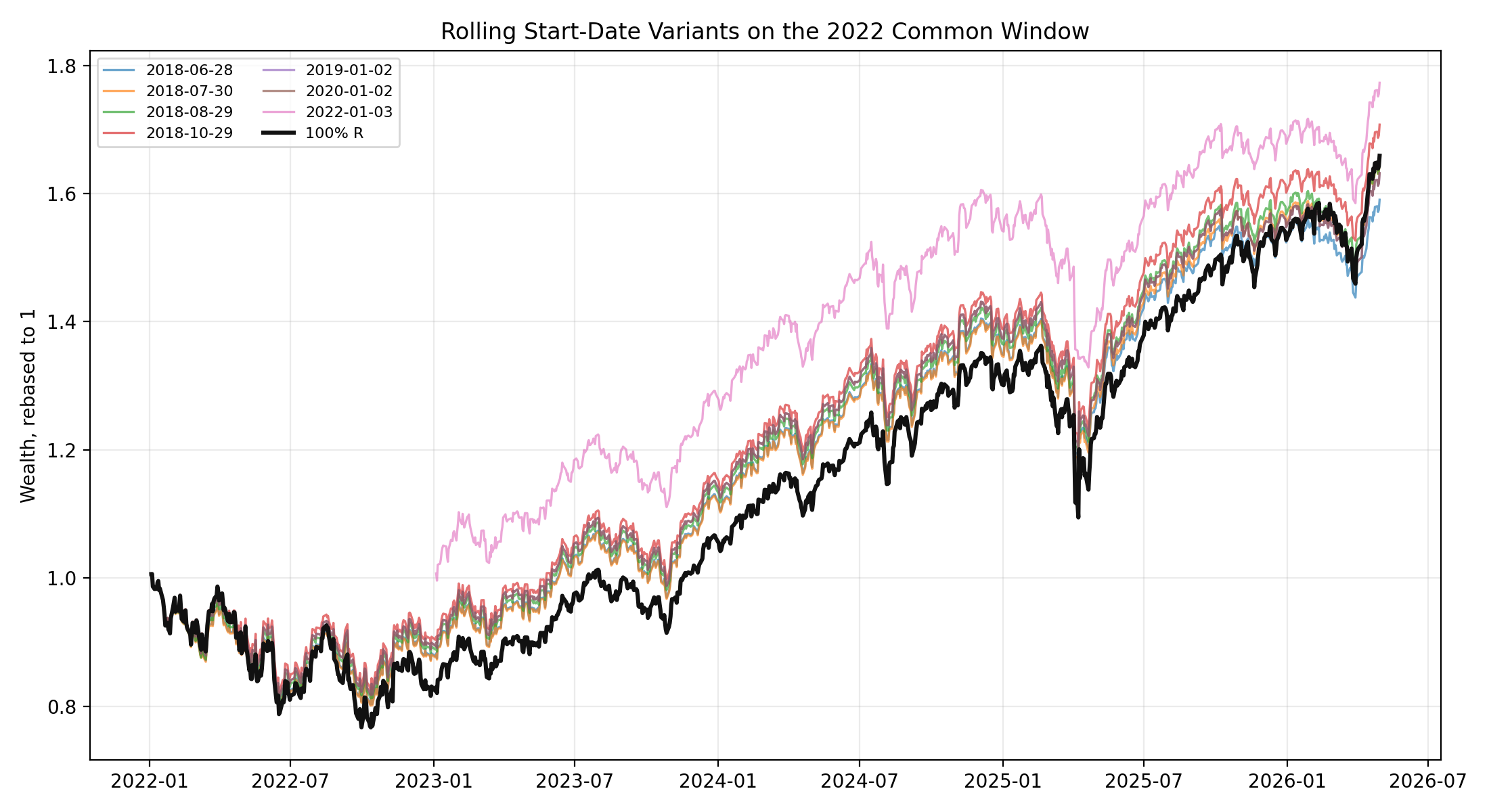}
\caption{Rolling Start-Date Variants on the Post-2022 Common Window}
\label{fig:rolling-common-equity}
\end{figure}

Rebasing all rolling requested-start variants on the same post-2022 window removes sample-length differences. In Figure~\ref{fig:rolling-common-equity}, most variants retain some drawdown control, but several lag the 100\% $R$ rebound. The common-window view therefore separates defensive usefulness from return dominance.

\begin{figure}[H]
\centering
\includegraphics[width=0.98\textwidth]{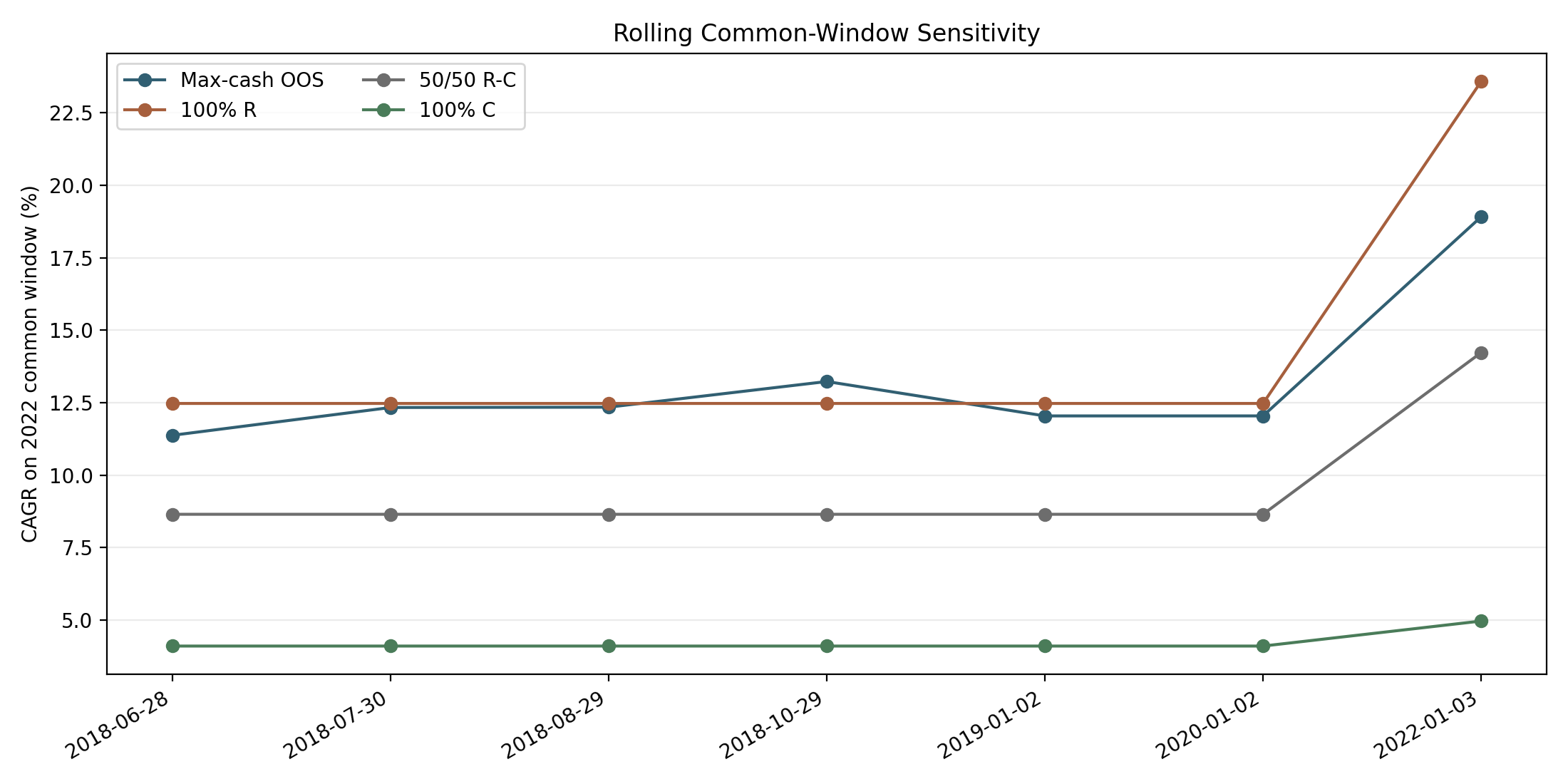}
\caption{Rolling Common-Window CAGR Sensitivity}
\label{fig:rolling-common-cagr}
\end{figure}

The same common-window rolling test can be summarized in CAGR terms. Figure~\ref{fig:rolling-common-cagr} shows that the current rolling specifications often fall below 100\% $R$ on post-2022 CAGR even when they reduce drawdown. This is a risk-management trade-off rather than a standalone alpha result.

\section{Expanding Start-Date Sensitivity}

The expanding version uses all available past data before each OOS block. This is the cleaner validation of a slowly accumulating empirical rule: every later block has a larger training sample, and the choice of requested start mainly changes the early OOS blocks and block alignment. In the current direct slow-tail version, expanding selection is stronger than rolling on the main window, but it no longer improves CAGR for every requested start. Its more stable result is drawdown reduction.

This does not remove model-selection risk, because the variable families and policy classes were constructed before the walk-forward exercise. It does show, however, that once the two module classes are fixed, the expanding parameter-selection process is not fragile to reasonable OOS start-date choices.

\begin{table}[H]
\centering
\caption{Expanding Start-Date Sensitivity: Start-to-End CAGR and Drawdown}
\label{tab:expanding-start-sens}
\begingroup
\tiny
\setlength{\tabcolsep}{3pt}
\resizebox{\textwidth}{!}{%
\begin{tabular}{lrrrrrrr}
\toprule
Requested Start & Max-Cash CAGR & Sharpe & Max DD & CAGR Diff & MDD Improvement & 100\% $R$ CAGR & 100\% $R$ Max DD\\
\midrule
2018-06-28 & \pct{19.35} & 1.13 & \pct{-22.05} & \pct{1.76} & \pct{11.53} & \pct{17.59} & \pct{-33.59}\\
2018-07-30 & \pct{18.72} & 1.10 & \pct{-22.05} & \pct{1.45} & \pct{11.53} & \pct{17.27} & \pct{-33.59}\\
2018-08-29 & \pct{19.94} & 1.16 & \pct{-22.05} & \pct{1.81} & \pct{11.53} & \pct{18.13} & \pct{-33.59}\\
2018-10-29 & \pct{19.29} & 1.12 & \pct{-22.05} & \pct{1.56} & \pct{11.53} & \pct{17.74} & \pct{-33.59}\\
2019-01-02 & \pct{18.13} & 1.07 & \pct{-22.05} & \pct{1.14} & \pct{11.53} & \pct{16.99} & \pct{-33.59}\\
2020-01-02 & \pct{15.79} & 1.05 & \pct{-19.98} & \pct{-0.05} & \pct{3.79} & \pct{15.84} & \pct{-23.78}\\
2022-01-03 & \pct{20.76} & 1.50 & \pct{-15.53} & \pct{-2.82} & \pct{4.13} & \pct{23.58} & \pct{-19.66}\\
\bottomrule
\end{tabular}
}
\endgroup
\end{table}

The expanding start-date result is stronger than the rolling counterpart on early requested starts. In Table~\ref{tab:expanding-start-sens}, requested starts through 2019 improve both CAGR and maximum drawdown relative to 100\% $R$. The 2020 and 2022 requested starts still reduce drawdown but trail on CAGR, which is consistent with the post-stress rebound cost documented in the component-level OOS tests.

\begin{table}[H]
\centering
\caption{Expanding Common-Window Sensitivity, 2022-01-03 to 2026-04-30}
\label{tab:expanding-common-sens}
\begingroup
\tiny
\setlength{\tabcolsep}{3pt}
\resizebox{\textwidth}{!}{%
\begin{tabular}{lrrrrrrr}
\toprule
Requested Start & Max-Cash CAGR & Max DD & 100\% $R$ CAGR & 50/50 $R-C$ CAGR & 100\% $C$ CAGR & CAGR Diff & MDD Improvement\\
\midrule
2018-06-28 & \pct{12.65} & \pct{-20.03} & \pct{12.47} & \pct{8.65} & \pct{4.11} & \pct{0.18} & \pct{3.75}\\
2018-07-30 & \pct{12.48} & \pct{-20.31} & \pct{12.47} & \pct{8.65} & \pct{4.11} & \pct{0.01} & \pct{3.47}\\
2018-08-29 & \pct{12.97} & \pct{-19.28} & \pct{12.47} & \pct{8.65} & \pct{4.11} & \pct{0.50} & \pct{4.49}\\
2018-10-29 & \pct{13.12} & \pct{-18.36} & \pct{12.47} & \pct{8.65} & \pct{4.11} & \pct{0.64} & \pct{5.41}\\
2019-01-02 & \pct{12.74} & \pct{-19.98} & \pct{12.47} & \pct{8.65} & \pct{4.11} & \pct{0.27} & \pct{3.79}\\
2020-01-02 & \pct{12.74} & \pct{-19.98} & \pct{12.47} & \pct{8.65} & \pct{4.11} & \pct{0.27} & \pct{3.79}\\
2022-01-03 & \pct{20.76} & \pct{-15.53} & \pct{23.58} & \pct{14.23} & \pct{4.97} & \pct{-2.82} & \pct{4.13}\\
\bottomrule
\end{tabular}
}
\endgroup
\end{table}

On the post-2022 common window, expanding selection is less sensitive to the requested start than rolling selection, but the result is still not a pure return improvement. Table~\ref{tab:expanding-common-sens} reports modest CAGR gains for pre-2022 requested starts and a material CAGR drag for the 2022 requested start, while every row improves maximum drawdown. This is the cleanest validation layer for the current defensive interpretation: the overlay buys drawdown reduction at the cost of some participation in the strongest rebound window.

\begin{figure}[H]
\centering
\includegraphics[width=0.98\textwidth]{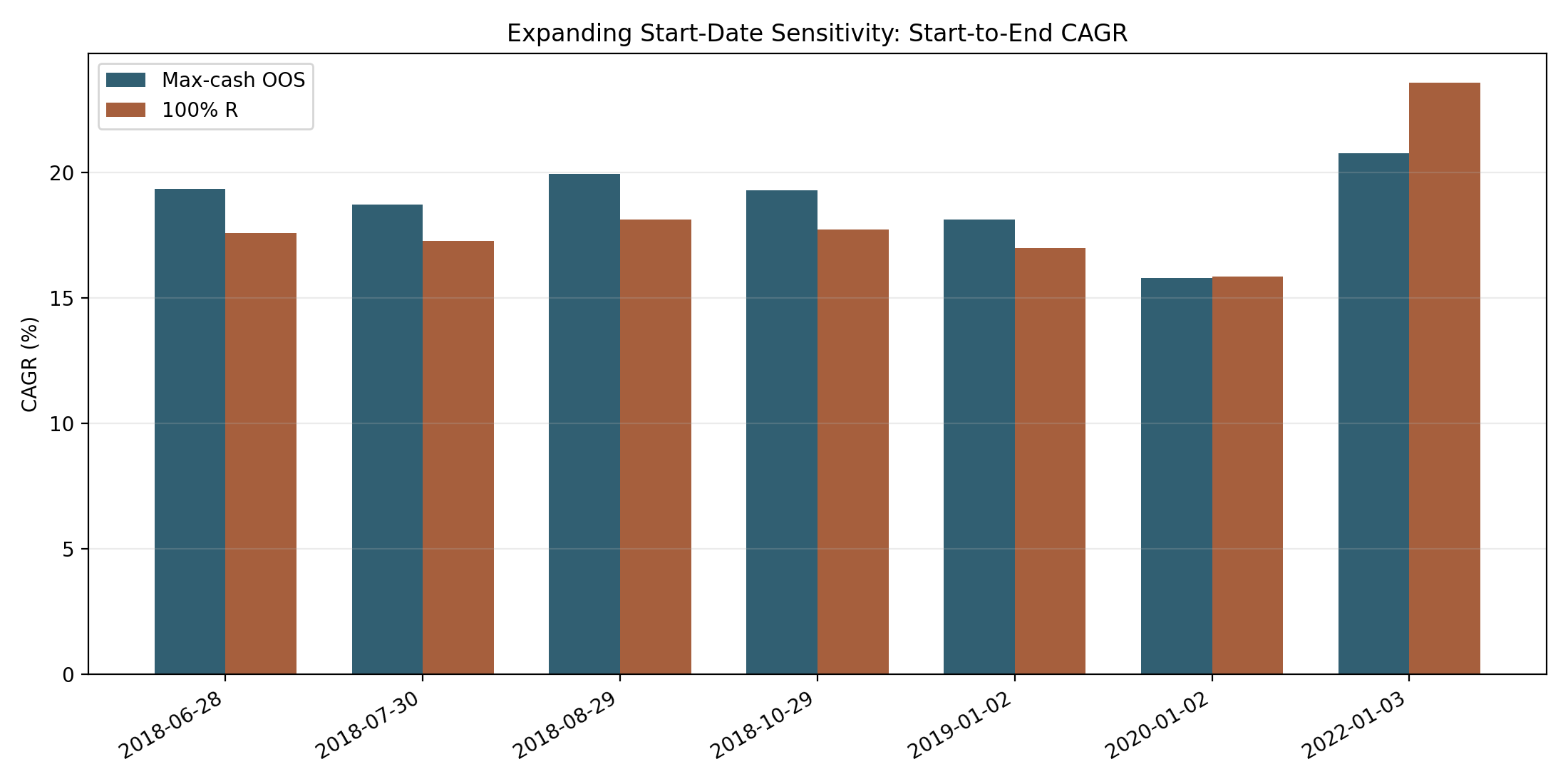}
\caption{Expanding Start-Date Sensitivity: Start-to-End CAGR}
\label{fig:expanding-start-bar}
\end{figure}

The stronger early-start result is also visible graphically. Expanding max-cash outperforms 100\% $R$ for requested starts before 2020, while later starts show the rebound-cost limitation (Figure~\ref{fig:expanding-start-bar}). The figure therefore supports a narrower claim than the earlier version: expanding selection is more stable than rolling, but it is still a defensive overlay.

\begin{figure}[H]
\centering
\includegraphics[width=0.98\textwidth]{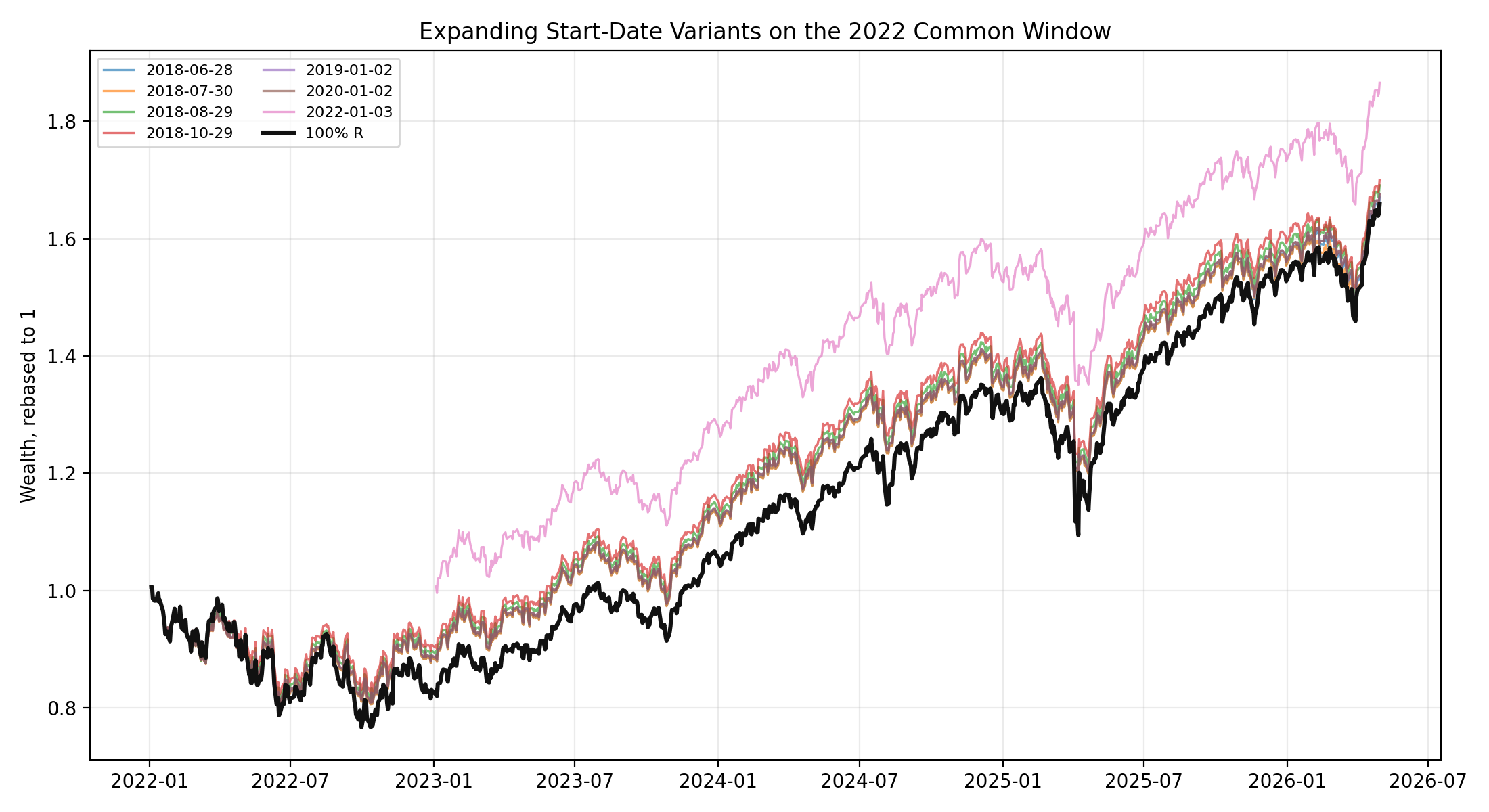}
\caption{Expanding Start-Date Variants on the Post-2022 Common Window}
\label{fig:expanding-common-equity}
\end{figure}

On the common post-2022 window, the expanding variants cluster more tightly than the rolling variants. This pattern in Figure~\ref{fig:expanding-common-equity} implies that the expanding result is not driven by one arbitrary requested start, but it also shows that the current direct slow-tail branch cannot dominate a strong risky-sleeve rebound on CAGR.

\begin{figure}[H]
\centering
\includegraphics[width=0.98\textwidth]{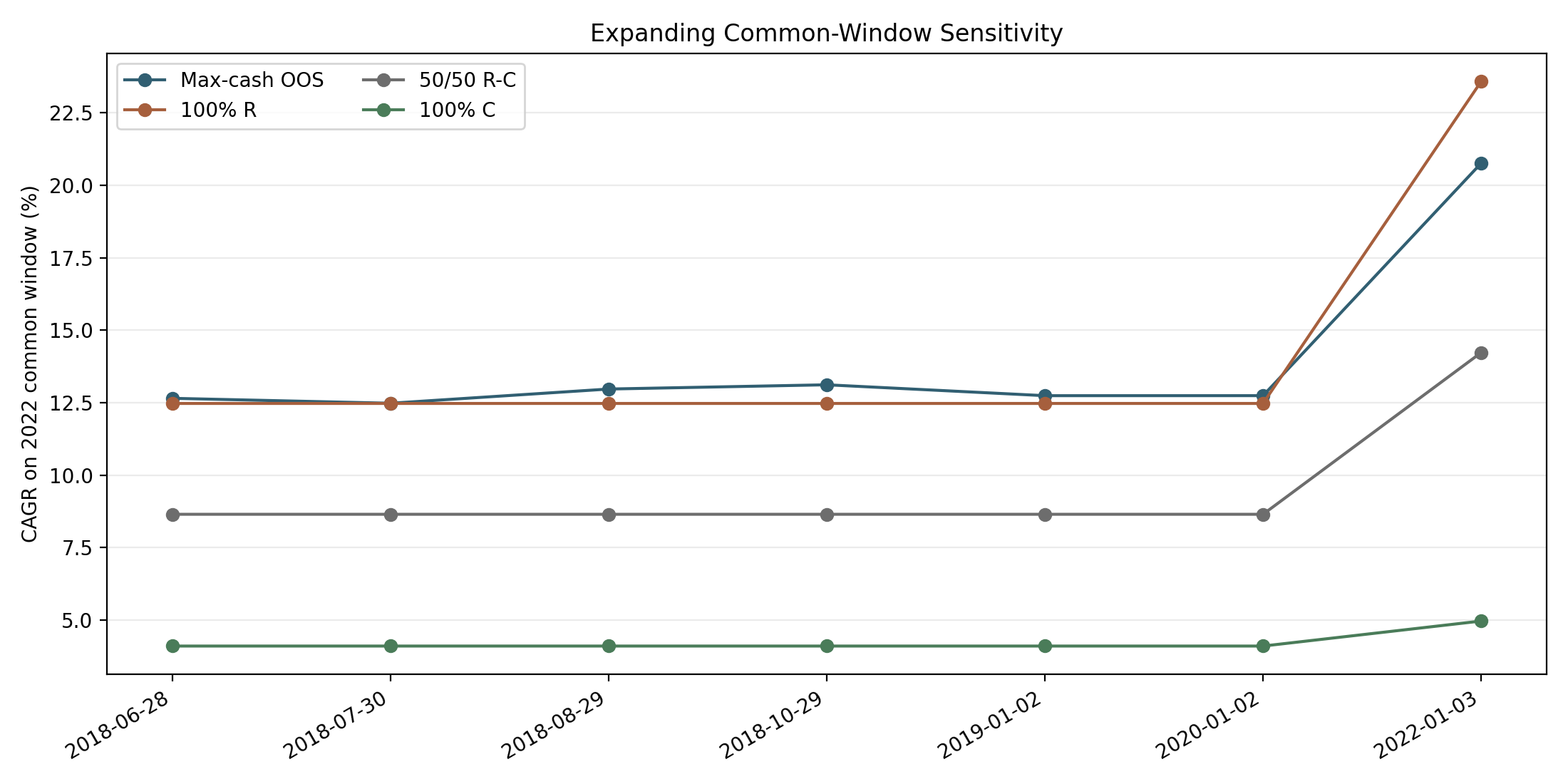}
\caption{Expanding Common-Window CAGR Sensitivity}
\label{fig:expanding-common-cagr}
\end{figure}

The common-window CAGR comparison gives the same conclusion in a more compact form. Expanding selection is close to or above 100\% $R$ for pre-2022 requested starts, but the 2022 requested start trails the rebound (Figure~\ref{fig:expanding-common-cagr}). This supports using expanding selection as the cleaner validation baseline for the cash-overlay setting, while still treating the overlay as a defensive tool rather than a guaranteed return enhancer.

\section{Economic Interpretation and Limitations}

The slow-tail filter and V-shape filter cover different risk shapes. The slow-tail filter is most useful when the risky sleeve is not sufficiently compensated over cash, as in 2022. The V-shape filter is most useful when fast stress creates a large drawdown risk, as in 2020 and April 2025. Their active cash-weight overlap is small, so the combined rule is not merely duplicating the same signal.

The economic value is mainly drawdown control with selective return improvement. The full-sample max-cash combination improves both CAGR and maximum drawdown. In walk-forward OOS tests, the main-window expanding and rolling combinations also improve both CAGR and maximum drawdown. Post-2022, however, both expanding and rolling lower drawdown but sacrifice CAGR. This is an important limitation of the current cash-overlay design: it can protect capital in stress, but it can trail a strong risky-sleeve rebound when cash exposure lingers.

The main limitation is data mining. Variables, interactions, and policy structures were developed on the available sample. The walk-forward tests re-select parameters from past data, but the current study does not yet perform real-time expanding re-screening of variables and interactions. It also does not yet report White's Reality Check, Hansen's SPA test, block-bootstrap inference, or the Deflated Sharpe Ratio. The evidence should therefore be read as a structured empirical strategy study, not as a final multiple-testing-adjusted anomaly claim.

\section{Conclusion}

This paper rewrites the cash-overlay experiment around a clean empirical object: a static equal-weight growth--defensive risky sleeve versus interest-bearing cash. The main methodological update is the direct slow-tail compensation module. It improves the 2022-style low-compensation regime less aggressively than the previous analogue construction, but with a simpler continuous score. The fixed V-shape module continues to reduce fast-crash losses in 2020 and 2025. The max-cash combination improves full-sample CAGR from 16.62\% to 18.83\% and improves maximum drawdown from -33.59\% to -18.05\%. In the stricter OOS-weight combination, the main expanding and rolling tests also improve both CAGR and drawdown relative to 100\% $R$.

The results support modular continuous cash overlays as economically meaningful risk-management tools. The next research step is to make variable selection itself real-time, add formal multiple-testing correction, and test the same design on longer or cross-asset histories.

\appendix
\section{Screening Tables and Robustness Details}

\subsection{Screening Protocol}

The screening layer is diagnostic rather than a direct trading rule. Each candidate main effect or interaction is evaluated against future $R-C$ returns at 21, 63, and 126 trading-day horizons. The 63-day horizon is the main economic horizon because it is long enough to capture a regime response but short enough to be relevant for tactical allocation. Since future labels overlap at daily frequency, ordinary OLS standard errors are not used as the primary inference object. The reported $t$-statistics use Newey--West/HAC standard errors.

For a candidate variable $x_t$, the basic diagnostic regression is
\begin{equation}
R^{R-C}_{t,t+h}=\alpha_h+\beta_h x_t+\epsilon_{t,h},\qquad h\in\{21,63,126\}.
\end{equation}
The coefficient sign, HAC $t$-value, non-overlapping 63-day coefficient, and multi-horizon direction are all reviewed. Interactions are included only when they have an economic interpretation. For example, a credit-panic term must combine market stress with credit deterioration; a low-compensation term must combine cash attractiveness, risky-sleeve strength, or low-volatility conditions that plausibly indicate limited incremental reward for staying fully in $R$.

The gate is intentionally weaker than a publication-grade multiple-testing claim. A variable can be retained as a component even when one diagnostic is mixed if the economic role is clear and the component improves portfolio behavior inside a continuous score. Conversely, a statistically strong variable is not automatically retained if it duplicates another signal, has poor coverage, or maps poorly into an implementable cash weight. This follows the workflow of the companion style-timing study \cite{xiong2026continuous}: statistical screens narrow the candidate set, while walk-forward policy performance determines whether the candidate survives as a tradable module.

\subsection{Slow-Tail Component Sets}

The current slow-tail module is not built around historical analogues. It uses a fixed set of 11 direct continuous components, grouped into cash-rising pressure, compressed risky-sleeve compensation, a direct rate headwind, and rate-path stress. The notation follows the construction in the main text: $CH$ is high cash yield, $CR$ is cash yield rising, $RH$ is rate headwind, $DD$ is drawdown depth, $VS$ is a VIX spike, $CRO$ is credit risk-off, $RS$ is risky-sleeve strength, $RU$ is risky-sleeve underwater, $LV$ is low VIX, and $LVOL$ is low risky-sleeve volatility. Every component listed here is continuous; after products are formed, they are expanding-standardized before entering the direct slow-tail score.

\begingroup
\scriptsize
\begin{longtable}{p{2.8cm}p{4.5cm}p{6.1cm}}
\caption{Direct Slow-Tail Components Used in the Policy Grid}
\label{tab:slow-component-sets}\\
\toprule
Block & Component Formula & Economic Meaning\\
\midrule
\endfirsthead
\toprule
Block & Component Formula & Economic Meaning\\
\midrule
\endhead
cash rising & $CR_t$ & Cash yield is rising, so the required compensation for holding $R$ over $C$ is higher.\\
compressed compensation & $RS_t \times LVOL_t$ & Risky sleeve has been strong in a low-volatility state; forward compensation may be compressed.\\
compressed compensation & $RS_t \times LVOL_t \times LV_t$ & Strong $R$, low realized volatility, and low VIX jointly describe complacent high-risk exposure.\\
compressed compensation & $RS_t \times LV_t$ & Risky-sleeve strength when volatility risk is priced cheaply.\\
compressed compensation & $CH_t \times RS_t \times LV_t$ & High cash yield plus strong risky sleeve plus low VIX; cash becomes competitive while risk appetite is still calm.\\
compressed compensation & $CH_t \times RS_t \times LVOL_t$ & High cash yield plus strong risky sleeve plus low realized volatility.\\
rate headwind & $RH_t$ & Rising 10-year yields as a direct rate headwind.\\
rate path & $RH_t \times VS_t$ & Rate headwind occurring with a VIX spike.\\
rate path & $RH_t \times CRO_t$ & Rate headwind occurring with credit risk-off.\\
rate path & $RH_t \times RU_t$ & Rate headwind while the risky sleeve is already underwater.\\
rate path & $DD_t \times RH_t \times CRO_t$ & Drawdown, rate headwind, and credit risk-off occurring together.\\
\bottomrule
\end{longtable}
\endgroup

The broader feature panel still contains legacy analogue columns for reproducibility, but they are not used by the current deployable slow-tail policy. The current score uses the component formulas listed in Table~\ref{tab:slow-component-sets}, the grid weights $\alpha_{comp}$, $\lambda_{rate}$, and $\lambda_{path}$, and the continuous cash mapping described in the main text.

\subsection{V-Shape Brake and Re-Entry Candidate Families}

The V-shape module is designed for fast drawdowns and fast recoveries. Its final material version uses the brake score as the trading score and keeps re-entry terms as diagnostics. The reason is practical: once the brake-derived raw cash weight falls below the 30\% material threshold, the policy begins returning to $R$ through the smoothing rule. A second hard re-entry trigger would add an additional threshold, which would conflict with the goal of keeping the final policy mostly continuous. In this table, $EZ(x)$ denotes the second expanding standardization applied after the primitive or product has been constructed.

\begingroup
\scriptsize
\begin{longtable}{p{2.8cm}p{5.5cm}p{5.5cm}}
\caption{V-Shape Candidate Families}
\label{tab:vshape-candidate-families}\\
\toprule
Family & Component Formula & Economic Meaning\\
\midrule
\endfirsthead
\toprule
Family & Component Formula & Economic Meaning\\
\midrule
\endhead
VIX brake core & $BrakeVIXLevel_t=EZ(VL_t)$ & High VIX level, where $VL_t=z(VIXPct_{756,t})$.\\
VIX brake core & $BrakeVIXSpike_t=EZ(VS10_t)$ & Fast VIX spike, where $VS10_t=z(\Delta VIX_{10,t})$.\\
Rate panic & $EZ(RR_t \times RL10_t \times VL_t)$ & Rate relief, risky-sleeve 10-day loss, and high VIX occurring together.\\
Rate panic & $EZ(RR_t \times CRO_t)$ & Rate relief occurring with credit risk-off.\\
Rate panic aggregate & $RatePanic_t=0.5EZ(RR_tRL10_tVL_t)+0.5EZ(RR_tCRO_t)$ & The score's rate-panic component.\\
Credit panic & $CreditPanic_t=EZ(VS10_t \times CW_t)$ & VIX spike with BAA--AAA spread widening.\\
Re-entry diagnostic & $ReentryOversold_t=EZ(DD_t \times VL_t)$ & Oversold risky sleeve with high VIX.\\
Re-entry diagnostic & $EZ(VR5_t \times CA5_t)$ & VIX relief with short-horizon credit appetite.\\
Re-entry diagnostic & $EZ(DD_t \times VR5_t)$ & Drawdown depth followed by VIX relief.\\
Re-entry diagnostic & $EZ(VL_t \times VR5_t \times CA5_t)$ & High VIX, VIX relief, and credit appetite together.\\
Re-entry diagnostic & $EZ(DD_t \times VR5_t \times CA5_t)$ & Deep drawdown, VIX relief, and credit appetite together.\\
Re-entry aggregate & Average of the four relief products, then $ReentryScore_t=0.8ReentryRelief_t+0.2ReentryOversold_t$. & Used for diagnostics only; it does not create an additional trading threshold.\\
\bottomrule
\end{longtable}
\endgroup

The primitive definitions are
\begin{align}
RR_t &= z(-\Delta TNX_{21,t}),&
RL10_t &= z(-RTrailing_{10,t}),\\
CRO_t &= z(-HYGSHYRel_{21,t}),&
CW_t &= z(\Delta CS_{21,t}),\\
DD_t &= z(-RDrawdown_t),&
VR5_t &= z(-\Delta VIX_{5,t}),\\
CA5_t &= z(HYGTrailing_{5,t}-SHYTrailing_{5,t}).
\end{align}
Thus the V-shape score uses falling yields only when they coincide with equity loss, volatility stress, or credit stress; ordinary rate relief by itself is not sufficient to create a large cash allocation.

\subsection{From Diagnostic Scores to Tradable Cash Weights}

Both modules convert continuous scores into continuous raw cash weights, then apply the same economic materiality rule. For slow-tail, the current direct score combines cash-rising pressure, compressed-risk-premium states, rate headwind, and rate-path stress:
\begin{equation}
CashRaw^{slow}_t
=MaxCash\cdot \sigma\left(\frac{SlowZ_t}{\tau}\right)^\gamma,
\end{equation}
where $\sigma(\cdot)$ is the logistic function and $\gamma$ is the cash convexity. For V-shape, the brake score is expanding-standardized and directly mapped into cash:
\begin{equation}
CashRaw^{v}_t
=MaxCash\cdot \sigma\left(\frac{BrakeZ_t}{\tau}\right)^\gamma.
\end{equation}
The shared rule is:
\begin{equation}
TargetCash_t=
\begin{cases}
CashRaw_t, & CashRaw_t\ge 0.30,\\
0, & CashRaw_t<0.30.
\end{cases}
\end{equation}
Entry to cash uses the target immediately; exit from cash uses the smoothing update
\begin{equation}
w^C_t=(1-\eta_{exit})w^C_{t-1}+\eta_{exit}TargetCash_t,\qquad \eta_{exit}=0.25.
\end{equation}
This is why the strategy can respond quickly during crash onset while avoiding excessive whipsaw when returning to the risky sleeve.

The first diagnostic layer is a screen, not a trading rule. The signs in Table~\ref{tab:main-screen} should not be read mechanically. Some stress variables have positive coefficients because high stress is often followed by recovery, while cash-rising and low-volatility variables point to weaker future $R-C$ compensation. This mixture is exactly why the paper separates slow-tail and V-shape modules instead of forcing all candidates into one monotone score.

\begin{table}[H]
\centering
\caption{Continuous Main-Effect Screen for Future $R-C$}
\label{tab:main-screen}
\tiny
\setlength{\tabcolsep}{2pt}
\begin{tabular}{p{2.9cm}p{2.1cm}rrrrrr}
\toprule
Variable & Role & 63d Coef. & 63d HAC $t$ & Nonoverlap Coef. & Nonoverlap $t$ & 21d HAC $t$ & 126d HAC $t$\\
\midrule
Cash yield high & cash compensation & -0.58 & -0.42 & 0.47 & 0.20 & -0.87 & -0.10\\
Cash yield rising & cash compensation & -1.73 & -2.43 & 1.29 & 0.47 & -1.82 & -1.20\\
Rate headwind & rate path & -1.61 & -1.26 & -5.79 & -2.78 & -0.28 & -1.63\\
Curve inversion & rate path & 0.32 & 0.31 & 1.76 & 0.94 & -0.21 & 0.21\\
Risky drawdown & risk stress & 2.92 & 2.30 & 3.52 & 2.27 & 2.62 & 1.82\\
High VIX & risk stress & 4.79 & 2.39 & 7.80 & 2.42 & 1.82 & 1.75\\
VIX spike & risk stress & 2.58 & 5.12 & 10.19 & 2.82 & 1.07 & 7.20\\
Credit risk-off & credit stress & 2.80 & 6.05 & 4.78 & 2.40 & 2.29 & 4.46\\
Credit widening & credit stress & 2.70 & 4.56 & 1.90 & 1.94 & 3.43 & 3.12\\
R strength & risk momentum & -1.91 & -0.90 & -1.58 & -0.48 & -0.93 & 0.28\\
R underwater & risk stress & 3.55 & 1.89 & 3.25 & 1.60 & 2.31 & 1.12\\
Low volatility & calm regime & -8.98 & -2.56 & -11.60 & -2.40 & -2.59 & -1.54\\
\bottomrule
\end{tabular}
\end{table}

The main-effect screen leaves two broad families. The first is a low-compensation family: cash yield rising, low volatility, and risky-sleeve strength in calm markets tend to imply lower forward $R-C$. The second is a stress-reversal family: VIX and credit-stress variables often have positive forward coefficients because severe stress is followed by rebound. The slow-tail screen in Table~\ref{tab:slow-screen} uses the first family, while the V-shape screen in Table~\ref{tab:v-screen} uses the second family as a crash-brake signal.

\begin{table}[H]
\centering
\caption{Slow-Tail Main and Interaction Component Screen}
\label{tab:slow-screen}
\tiny
\setlength{\tabcolsep}{2pt}
\begin{tabular}{p{4.1cm}p{1.8cm}rrrrrr}
\toprule
Component & Role & 63d Coef. & 63d HAC $t$ & Nonoverlap Coef. & Nonoverlap $t$ & 21d HAC $t$ & 126d HAC $t$\\
\midrule
Cash rising & main effect & -0.25 & -3.35 & 0.70 & 0.29 & -2.74 & -4.82\\
R strength x low vol & two-way & 2.62 & 3.81 & 2.31 & 1.07 & 4.73 & 4.49\\
R strength x low vol x low VIX & three-way & -3.29 & -7.06 & -3.88 & -1.77 & -4.55 & -7.40\\
R strength x low VIX & two-way & 1.55 & 0.97 & 1.74 & 0.97 & 2.75 & 0.51\\
Cash high x R strength x low VIX & three-way & 0.33 & 0.35 & 2.58 & 1.63 & 0.35 & 0.85\\
Cash high x R strength x low vol & three-way & 0.05 & 0.07 & 4.67 & 1.96 & 0.23 & 0.22\\
Rate headwind & main effect & -0.66 & -0.75 & -3.68 & -2.22 & 0.47 & -1.22\\
Rate headwind x VIX spike & two-way & -1.04 & -1.41 & -0.87 & -0.44 & 0.82 & -1.22\\
Rate headwind x credit risk-off & two-way & -1.96 & -2.75 & -3.39 & -1.27 & -0.83 & -2.61\\
Rate headwind x R underwater & two-way & -2.60 & -2.09 & -2.75 & -1.52 & -2.24 & -2.84\\
Drawdown x rate headwind x credit risk-off & three-way & -2.52 & -7.97 & -6.76 & -2.28 & -1.59 & -7.11\\
Slow direct score & aggregate score & -0.39 & -3.50 & -3.22 & -1.40 & -2.92 & -3.30\\
\bottomrule
\end{tabular}
\end{table}

The direct slow-tail score survives because the combined state is more informative than many isolated interactions. In Table~\ref{tab:slow-screen}, the aggregate score has a negative 63-day coefficient and a negative non-overlapping coefficient, meaning worse direct slow-tail states correspond to weaker future $R-C$ compensation. This aggregate evidence is more important than any single component because the policy trades on the combined compensation score, not on isolated interaction terms.

\begin{table}[H]
\centering
\caption{V-Shape Brake and Re-Entry Component Screen}
\label{tab:v-screen}
\tiny
\setlength{\tabcolsep}{2pt}
\begin{tabular}{p{3.0cm}p{2.5cm}rrrrr}
\toprule
Component & Role & 63d Coef. & 63d HAC $t$ & Nonoverlap Coef. & Nonoverlap $t$ & 21d HAC $t$\\
\midrule
VIX level & brake main & 1.37 & 1.73 & 4.89 & 2.68 & 1.49\\
VIX spike & brake main & 0.50 & 1.41 & -2.83 & -1.70 & -0.06\\
Rate panic & brake interaction & 1.87 & 3.35 & -5.57 & -1.17 & 0.03\\
Credit panic & brake interaction & 0.90 & 2.21 & -2.95 & -0.97 & 0.39\\
Reentry oversold & reentry diagnostic & 1.92 & 1.07 & 8.73 & 2.69 & 1.66\\
Reentry relief & reentry diagnostic & 2.66 & 2.93 & 10.80 & 2.17 & 0.41\\
\bottomrule
\end{tabular}
\end{table}

The V-shape screen requires a different interpretation from the slow-tail screen. Table~\ref{tab:v-screen} is not a simple ``high score means low future return'' table. For crash brakes, high stress can precede positive forward $R-C$ because the crash has already occurred and recovery follows. The brake score is therefore evaluated by event loss reduction and drawdown control, not only by the sign of a 63-day predictive coefficient. The re-entry rows are kept as diagnostics because they describe rebound conditions but are not used as an additional hard trading trigger in the final material policy.

\begin{table}[H]
\centering
\caption{Slow-Tail Entry and Active-State Diagnostics}
\label{tab:slow-entry}
\begingroup
\tiny
\setlength{\tabcolsep}{2pt}
\resizebox{\textwidth}{!}{%
\begin{tabular}{lrrrrrrr}
\toprule
Bucket & Days & Future 21d $R-C$ & Future 63d $R-C$ & Future 126d $R-C$ & 63d Win Rate & Avg. Raw Cash & Avg. Cash\\
\midrule
entry days & 17 & \pct{-0.94} & \pct{-1.45} & \pct{-1.39} & \pct{41.18} & \pct{31.79} & \pct{8.29}\\
material active days & 93 & \pct{-2.33} & \pct{-2.09} & \pct{-0.15} & \pct{38.71} & \pct{34.26} & \pct{30.15}\\
cash raw above 30pct & 92 & \pct{-2.25} & \pct{-2.11} & \pct{-0.17} & \pct{38.04} & \pct{34.31} & \pct{30.00}\\
actual cash weight days & 202 & \pct{-0.27} & \pct{0.00} & \pct{2.35} & \pct{49.50} & \pct{30.09} & \pct{21.39}\\
\bottomrule
\end{tabular}
}
\endgroup
\end{table}

Slow-tail activation is economically sparse. Table~\ref{tab:slow-entry} reports 17 entry days and 93 material active days in the main window. The future 63-day $R-C$ mean is negative for the material active state, which confirms that the cash allocation is concentrated in the intended weak-compensation regime. The broader actual-cash-weight row includes smoothed exits, so its future $R-C$ average is less negative; this is expected because recovery days are included during the exit path.

\begin{table}[H]
\centering
\caption{Slow-Tail Cost Sensitivity}
\label{tab:slow-cost}
\scriptsize
\begin{tabular}{rrrrrr}
\toprule
Cost bp & CAGR & Sharpe & Max DD & Ann. Turnover & Avg. Cash\\
\midrule
0 & \pct{17.96} & 0.94 & \pct{-33.59} & \pct{108.36} & \pct{2.19}\\
5 & \pct{17.89} & 0.94 & \pct{-33.59} & \pct{108.36} & \pct{2.19}\\
10 & \pct{17.83} & 0.94 & \pct{-33.59} & \pct{108.36} & \pct{2.19}\\
20 & \pct{17.70} & 0.93 & \pct{-33.59} & \pct{108.36} & \pct{2.19}\\
\bottomrule
\end{tabular}
\end{table}

Transaction costs do not materially alter the slow-tail conclusion over the tested range. As reported in Table~\ref{tab:slow-cost}, annual turnover is modest compared with the V-shape module, so moving from 0bp to 20bp reduces CAGR only gradually. This matters because the slow-tail module is meant to handle multi-week or multi-month deterioration, not high-frequency switching.

\begin{table}[H]
\centering
\caption{Slow-Tail Fixed Holdout Check}
\label{tab:slow-hold}
\begingroup
\tiny
\setlength{\tabcolsep}{2pt}
\resizebox{\textwidth}{!}{%
\begin{tabular}{llrrrrrr}
\toprule
Period & Role & CAGR & Max DD & Ann. Turnover & Avg. Cash & 100\% $R$ CAGR & 100\% $R$ Max DD\\
\midrule
2018--2021 train & fixed direct\_slow\_157 & \pct{21.98} & \pct{-33.59} & \pct{17.12} & \pct{0.13} & \pct{22.08} & \pct{-33.59}\\
2022--2026 holdout & fixed direct\_slow\_157 & \pct{13.72} & \pct{-19.66} & \pct{195.54} & \pct{4.51} & \pct{12.47} & \pct{-23.78}\\
\bottomrule
\end{tabular}
}
\endgroup
\end{table}

The fixed train/holdout check is deliberately simple. In Table~\ref{tab:slow-hold}, a rule selected on the 2018--2021 training period is almost inactive during that training period, but it improves both CAGR and drawdown in the 2022--2026 holdout window. This supports the interpretation that the slow-tail state becomes valuable only once cash yield and risky-sleeve compensation conditions change after 2021.

\begin{table}[H]
\centering
\caption{V-Shape Score Sorting Diagnostics}
\label{tab:v-score}
\scriptsize
\begin{tabular}{llrrr}
\toprule
Score & Bucket & Days & Future 63d $R-C$ Mean & 63d Win Rate\\
\midrule
brake z & Q1 & 430 & \pct{2.31} & \pct{68.37}\\
brake z & Q2 & 429 & \pct{2.69} & \pct{75.06}\\
brake z & Q3 & 429 & \pct{2.80} & \pct{71.33}\\
brake z & Q4 & 429 & \pct{5.04} & \pct{80.89}\\
brake z & Q5 & 429 & \pct{4.19} & \pct{71.33}\\
reentry z & Q1 & 431 & \pct{4.46} & \pct{81.67}\\
reentry z & Q2 & 430 & \pct{2.79} & \pct{72.09}\\
reentry z & Q3 & 430 & \pct{2.25} & \pct{68.37}\\
reentry z & Q4 & 430 & \pct{1.96} & \pct{64.42}\\
reentry z & Q5 & 430 & \pct{5.58} & \pct{80.70}\\
\bottomrule
\end{tabular}
\end{table}

The score-sorting exercise is diagnostic rather than a portfolio rule. The brake score does not need monotone Q1--Q5 return sorting because it is not intended to forecast ordinary returns across all states (Table~\ref{tab:v-score}). Its purpose is to identify high-stress tails where temporary cash has drawdown value. The re-entry score has a high Q5 future mean, which is consistent with rebound conditions, but it remains diagnostic to avoid adding a second discretionary threshold.

\begin{table}[H]
\centering
\caption{V-Shape Cost Sensitivity}
\label{tab:v-cost}
\scriptsize
\begin{tabular}{rrrrrr}
\toprule
Cost bp & CAGR & Sharpe & Max DD & Ann. Turnover & Avg. Cash\\
\midrule
0 & \pct{17.94} & 1.10 & \pct{-23.77} & \pct{532.41} & \pct{3.65}\\
5 & \pct{17.63} & 1.08 & \pct{-23.77} & \pct{532.41} & \pct{3.65}\\
10 & \pct{17.32} & 1.07 & \pct{-23.77} & \pct{532.41} & \pct{3.65}\\
20 & \pct{16.69} & 1.03 & \pct{-23.78} & \pct{532.41} & \pct{3.65}\\
\bottomrule
\end{tabular}
\end{table}

The V-shape module is more cost-sensitive because it trades more. Table~\ref{tab:v-cost} shows a stronger cost effect than Table~\ref{tab:slow-cost}. Even at 20bp, however, maximum drawdown remains materially below 100\% $R$. This reinforces the interpretation that V-shape is a risk-control component whose economic case comes primarily from drawdown reduction.

\begin{table}[H]
\centering
\caption{V-Shape Yearly Breakdown Versus 100\% $R$}
\label{tab:v-year}
\tiny
\setlength{\tabcolsep}{3pt}
\begin{tabular}{rrrrrr}
\toprule
Year & V-shape CAGR & V-shape Max DD & Avg. Cash & 100\% $R$ CAGR & 100\% $R$ Max DD\\
\midrule
2017 & \pct{23.08} & \pct{-1.88} & \pct{5.59} & \pct{27.03} & \pct{-2.08}\\
2018 & \pct{-2.96} & \pct{-17.00} & \pct{7.68} & \pct{-3.03} & \pct{-19.71}\\
2019 & \pct{32.32} & \pct{-8.03} & \pct{0.57} & \pct{32.90} & \pct{-8.03}\\
2020 & \pct{43.96} & \pct{-15.19} & \pct{7.64} & \pct{23.71} & \pct{-33.59}\\
2021 & \pct{29.24} & \pct{-4.92} & \pct{0.84} & \pct{31.04} & \pct{-4.92}\\
2022 & \pct{-17.12} & \pct{-23.77} & \pct{0.00} & \pct{-17.12} & \pct{-23.78}\\
2023 & \pct{28.40} & \pct{-9.73} & \pct{0.00} & \pct{28.40} & \pct{-9.73}\\
2024 & \pct{18.21} & \pct{-8.00} & \pct{3.28} & \pct{22.51} & \pct{-8.88}\\
2025 & \pct{15.37} & \pct{-14.82} & \pct{7.02} & \pct{17.46} & \pct{-19.66}\\
2026 & \pct{20.94} & \pct{-8.02} & \pct{7.61} & \pct{28.88} & \pct{-7.95}\\
\bottomrule
\end{tabular}
\end{table}

The yearly decomposition clarifies where the V-shape filter earns its keep. In Table~\ref{tab:v-year}, the module's major positive contribution is 2020, where it reduces the crash drawdown and then still participates enough in the rebound to outperform the risky sleeve for the year. It does not help in 2022 because the 2022 regime is not a fast V-shaped crash. In 2024--2026, the module gives up some return in exchange for lower drawdowns during stress episodes.

\begin{table}[H]
\centering
\caption{Complete OOS Return Matrix from Sub-Strategy OOS Weights}
\label{tab:oos-complete}
\tiny
\setlength{\tabcolsep}{3pt}
\begin{tabular}{lllrrrrrr}
\toprule
OOS Window & Mode & Method & CAGR & Max DD & Ann. Turnover & Avg. Cash & CAGR Diff & MDD Improvement\\
\midrule
main-oos & Expanding & Max-cash & \pct{19.35} & \pct{-22.05} & \pct{305.09} & \pct{5.22} & \pct{1.76} & \pct{11.53}\\
main-oos & Expanding & Slow-tail & \pct{18.79} & \pct{-33.59} & \pct{114.57} & \pct{2.44} & \pct{1.20} & \pct{0.00}\\
main-oos & Expanding & V-shape & \pct{17.98} & \pct{-25.82} & \pct{202.58} & \pct{2.92} & \pct{0.39} & \pct{7.77}\\
main-oos & Expanding & 100\% $R$ & \pct{17.59} & \pct{-33.59} & \pct{0.00} & \pct{0.00} & \pct{0.00} & \pct{0.00}\\
main-oos & Rolling & Max-cash & \pct{18.50} & \pct{-22.05} & \pct{351.81} & \pct{6.15} & \pct{0.90} & \pct{11.53}\\
main-oos & Rolling & Slow-tail & \pct{18.12} & \pct{-33.59} & \pct{138.31} & \pct{2.77} & \pct{0.53} & \pct{0.00}\\
main-oos & Rolling & V-shape & \pct{17.22} & \pct{-26.18} & \pct{246.39} & \pct{3.73} & \pct{-0.37} & \pct{7.40}\\
main-oos & Rolling & 100\% $R$ & \pct{17.59} & \pct{-33.59} & \pct{0.00} & \pct{0.00} & \pct{0.00} & \pct{0.00}\\
post2022-oos & Expanding & Max-cash & \pct{20.76} & \pct{-15.53} & \pct{397.21} & \pct{4.56} & \pct{-2.82} & \pct{4.13}\\
post2022-oos & Expanding & Slow-tail & \pct{21.19} & \pct{-19.74} & \pct{169.65} & \pct{1.62} & \pct{-2.39} & \pct{-0.08}\\
post2022-oos & Expanding & V-shape & \pct{21.90} & \pct{-14.75} & \pct{274.19} & \pct{3.38} & \pct{-1.67} & \pct{4.92}\\
post2022-oos & Expanding & 100\% $R$ & \pct{23.58} & \pct{-19.66} & \pct{0.00} & \pct{0.00} & \pct{0.00} & \pct{0.00}\\
post2022-oos & Rolling & Max-cash & \pct{18.91} & \pct{-17.26} & \pct{470.06} & \pct{6.24} & \pct{-4.67} & \pct{2.41}\\
post2022-oos & Rolling & Slow-tail & \pct{21.19} & \pct{-19.74} & \pct{169.65} & \pct{1.62} & \pct{-2.39} & \pct{-0.08}\\
post2022-oos & Rolling & V-shape & \pct{19.35} & \pct{-16.88} & \pct{370.97} & \pct{5.24} & \pct{-4.23} & \pct{2.78}\\
post2022-oos & Rolling & 100\% $R$ & \pct{23.58} & \pct{-19.66} & \pct{0.00} & \pct{0.00} & \pct{0.00} & \pct{0.00}\\
\bottomrule
\end{tabular}
\end{table}

The most compact audit of the combined walk-forward experiment is the full OOS return matrix. Table~\ref{tab:oos-complete} reports component returns and combined returns on the same OOS windows. The main OOS window supports the combination: max-cash improves both CAGR and drawdown relative to 100\% $R$. The post-2022 rows show the limitation of a simple upper-envelope rule: it reduces drawdown relative to 100\% $R$, but it can underperform the best standalone component and the risky sleeve on CAGR when the rebound is strong. The comparison clarifies the combination objective: improve defensive balance, not maximize standalone component CAGR in every subperiod.

\begin{table}[H]
\centering
\caption{Walk-Forward Parameter Selection Diagnostics}
\label{tab:wf-selection}
\begingroup
\tiny
\setlength{\tabcolsep}{3pt}
\resizebox{\textwidth}{!}{%
\begin{tabular}{lllrrlrrr}
\toprule
Module & OOS Window & Mode & Blocks & Unique Configs & Test Window & Top Config Blocks & Top Share & Median Train CAGR\\
\midrule
Slow-tail & main-oos & Expanding & 28 & 4 & 2019-07-01--2026-04-30 & 16 & \pct{57.14} & --\\
Slow-tail & main-oos & Rolling & 28 & 10 & 2019-07-01--2026-04-30 & 7 & \pct{25.00} & --\\
Slow-tail & post2022-oos & Expanding & 14 & 6 & 2023-01-04--2026-04-30 & 5 & \pct{35.71} & --\\
Slow-tail & post2022-oos & Rolling & 14 & 6 & 2023-01-04--2026-04-30 & 5 & \pct{35.71} & --\\
V-shape & main-oos & Expanding & 31 & 8 & 2018-09-27--2026-04-30 & 14 & \pct{45.16} & \pct{17.22}\\
V-shape & main-oos & Rolling & 31 & 20 & 2018-09-27--2026-04-30 & 4 & \pct{12.90} & \pct{16.89}\\
V-shape & post2022-oos & Expanding & 18 & 4 & 2022-01-03--2026-04-30 & 14 & \pct{77.78} & \pct{17.14}\\
V-shape & post2022-oos & Rolling & 18 & 13 & 2022-01-03--2026-04-30 & 5 & \pct{27.78} & \pct{17.52}\\
\bottomrule
\end{tabular}
}
\endgroup
\end{table}

Parameter re-selection is substantial but structured. Table~\ref{tab:wf-selection} shows that expanding selection concentrates more heavily in a few top configurations, while rolling selection uses more unique configurations, especially for V-shape. The direct slow-tail selection files record the chosen configuration by block but do not store the training-period performance metrics, so the median training CAGR cells are left blank for that module. The configuration-count evidence is still useful: slow-tail is more concentrated under expanding selection, whereas V-shape remains deliberately more adaptive.

\section{Start-Date Sensitivity Selection Stability}

Portfolio performance and parameter stability tell different parts of the validation story. The performance tables above summarize realized results across requested starts, while Table~\ref{tab:start-selection-stability} reports the parameter-selection layer behind those results. The slow-tail module is more stable under expanding selection, while the V-shape module is deliberately more adaptive under rolling selection. This is expected: the V-shape filter is a crash brake whose preferred VIX, rate, and credit-panic weights can shift after recent fast-crash episodes. As in Table~\ref{tab:wf-selection}, the direct slow-tail selection artifacts record block-level configuration choices but not training-period performance columns, so those training-metric cells are intentionally left blank.

\begingroup
\tiny
\setlength{\tabcolsep}{2pt}
\begin{longtable}{p{1.65cm}p{1.85cm}p{1.35cm}rrrrr}
\caption{Start-Date Sensitivity: Component Selection Stability Across Requested Starts}
\label{tab:start-selection-stability}\\
\toprule
Component & Start & Mode & Blocks & Unique & Top Share & Med. Train CAGR & Med. Train Sharpe\\
\midrule
\endfirsthead
\toprule
Component & Start & Mode & Blocks & Unique & Top Share & Med. Train CAGR & Med. Train Sharpe\\
\midrule
\endhead
Slow-tail & 2018-06-28 & Expanding & 28 & 4 & \pct{57.14} & -- & --\\
V-shape & 2018-06-28 & Expanding & 31 & 8 & \pct{45.16} & \pct{17.22} & 1.05\\
Slow-tail & 2018-06-28 & Rolling & 28 & 8 & \pct{25.00} & -- & --\\
V-shape & 2018-06-28 & Rolling & 31 & 20 & \pct{12.90} & \pct{16.89} & 1.02\\
Slow-tail & 2018-07-30 & Expanding & 27 & 5 & \pct{59.26} & -- & --\\
V-shape & 2018-07-30 & Expanding & 31 & 9 & \pct{45.16} & \pct{17.38} & 1.06\\
Slow-tail & 2018-07-30 & Rolling & 27 & 11 & \pct{37.04} & -- & --\\
V-shape & 2018-07-30 & Rolling & 31 & 21 & \pct{12.90} & \pct{16.67} & 1.09\\
Slow-tail & 2018-08-29 & Expanding & 27 & 4 & \pct{62.96} & -- & --\\
V-shape & 2018-08-29 & Expanding & 31 & 9 & \pct{41.94} & \pct{17.21} & 1.04\\
Slow-tail & 2018-08-29 & Rolling & 27 & 9 & \pct{37.04} & -- & --\\
V-shape & 2018-08-29 & Rolling & 31 & 18 & \pct{12.90} & \pct{16.11} & 1.04\\
Slow-tail & 2018-10-29 & Expanding & 26 & 4 & \pct{65.38} & -- & --\\
V-shape & 2018-10-29 & Expanding & 30 & 7 & \pct{50.00} & \pct{17.24} & 1.05\\
Slow-tail & 2018-10-29 & Rolling & 26 & 9 & \pct{42.31} & -- & --\\
V-shape & 2018-10-29 & Rolling & 30 & 19 & \pct{16.67} & \pct{16.56} & 1.04\\
Slow-tail & 2019-01-02 & Expanding & 26 & 4 & \pct{65.38} & -- & --\\
V-shape & 2019-01-02 & Expanding & 30 & 6 & \pct{46.67} & \pct{17.06} & 1.05\\
Slow-tail & 2019-01-02 & Rolling & 26 & 8 & \pct{34.62} & -- & --\\
V-shape & 2019-01-02 & Rolling & 30 & 18 & \pct{13.33} & \pct{16.85} & 1.09\\
Slow-tail & 2020-01-02 & Expanding & 22 & 3 & \pct{77.27} & -- & --\\
V-shape & 2020-01-02 & Expanding & 26 & 5 & \pct{53.85} & \pct{17.36} & 1.05\\
Slow-tail & 2020-01-02 & Rolling & 22 & 8 & \pct{40.91} & -- & --\\
V-shape & 2020-01-02 & Rolling & 26 & 17 & \pct{15.38} & \pct{18.24} & 1.15\\
Slow-tail & 2022-01-03 & Expanding & 14 & 6 & \pct{35.71} & -- & --\\
V-shape & 2022-01-03 & Expanding & 18 & 4 & \pct{77.78} & \pct{17.14} & 1.01\\
Slow-tail & 2022-01-03 & Rolling & 14 & 7 & \pct{35.71} & -- & --\\
V-shape & 2022-01-03 & Rolling & 18 & 13 & \pct{27.78} & \pct{17.52} & 0.94\\
\bottomrule
\end{longtable}
\endgroup

The seven requested starts tell the same selection-stability story. In Table~\ref{tab:start-selection-stability}, the slow-tail expanding selector is relatively concentrated, especially for pre-2022 requested starts, while V-shape rolling selection remains dispersed across many configurations. This supports the paper's interpretation of the two modules: slow-tail is a persistent-regime filter, whereas V-shape is a more event-sensitive brake.


\begin{thebibliography}{99}

\bibitem[Bailey and Lopez de Prado(2014)]{bailey2014probability}
Bailey, D. H., and M. Lopez de Prado. 2014.
The Deflated Sharpe Ratio: Correcting for selection bias, backtest overfitting, and non-normality.
\emph{Journal of Portfolio Management} 40(5): 94--107.

\bibitem[Brandt, Santa-Clara, and Valkanov(2009)]{brandt2009parametric}
Brandt, M. W., P. Santa-Clara, and R. Valkanov. 2009.
Parametric portfolio policies: Exploiting characteristics in the cross-section of equity returns.
\emph{Review of Financial Studies} 22(9): 3411--3447.

\bibitem[Campbell and Thompson(2008)]{campbell2008predicting}
Campbell, J. Y., and S. B. Thompson. 2008.
Predicting excess stock returns out of sample: Can anything beat the historical average?
\emph{Review of Financial Studies} 21(4): 1509--1531.

\bibitem[Goyal and Welch(2008)]{goyal2008comprehensive}
Goyal, A., and I. Welch. 2008.
A comprehensive look at the empirical performance of equity premium prediction.
\emph{Review of Financial Studies} 21(4): 1455--1508.

\bibitem[Hansen(2005)]{hansen2005superior}
Hansen, P. R. 2005.
A test for superior predictive ability.
\emph{Journal of Business and Economic Statistics} 23(4): 365--380.

\bibitem[Xiong(2026)]{xiong2026continuous}
Xiong, Z. 2026.
Continuous timing signals for growth--defensive style allocation: Factor attribution, risk matching, and out-of-sample evidence.
Working paper.

\bibitem[White(2000)]{white2000reality}
White, H. 2000.
A reality check for data snooping.
\emph{Econometrica} 68(5): 1097--1126.

\end{thebibliography}
\end{document}